\documentclass[11pt]{article}
\usepackage{a4wide}
\usepackage{amsfonts,amsmath}
\usepackage[english]{babel}
\usepackage{palatino}
\usepackage{graphicx}
\newcommand{\vq}{\vec{q}}
\newcommand{\vk}{\vec{k}}
\newcommand{\ph}{\vert\psi_p\rangle}

\newcommand{\MSbar}{\overline{\mbox{MS}}}

\newcommand{\al}{\alpha}

\newcommand{\p}{\partial}
\newcommand{\oc}{\overline{c}}

\newcommand{\wm}{\widetilde{m}}

\newcommand{\oB}{\overline{B}}
\newcommand{\occ}{\overline{c}}
\newcommand{\oG}{\overline{G}}
\newcommand{\be}{\beta}  
\newcommand{\sect}[1]{ \section{#1} \setcounter{equation}{0} }

\begin{document}
\date{}
\title{\textbf{Unitarity analysis of a non-Abelian gauge invariant action with a mass}}
\author{\textbf{D. Dudal}\thanks{david.dudal@ugent.be}, \textbf{N.
Vandersickel}\thanks{nele.vandersickel@ugent.be}, \textbf{H.
Verschelde}\thanks{henri.verschelde@ugent.be}\\\\
\textit{{\small Ghent University}} \\
\textit{{\small Department of Mathematical Physics and Astronomy}} \\
\textit{{\small Krijgslaan 281-S9, B-9000 Gent, Belgium}}}
\maketitle
\begin{abstract}
\noindent In previous work done by us and coworkers, we have been
able to construct a local, non-Abelian gauge invariant action with a
mass parameter, based on the nonlocal gauge invariant mass dimension
two operator $F_{\mu\nu} (D^2)^{-1} F_{\mu\nu}$. The
renormalizability of the resulting action was proven to all orders
of perturbation theory, in the class of linear covariant gauges. We
also discussed the perturbative equivalence of the model with
ordinary massless Yang-Mills gauge theories when the mass is
identically zero. Furthermore, we pointed out the existence of a
BRST symmetry with corresponding nilpotent charge. In this paper, we
study the issue of unitarity of this massive gauge model. Firstly,
we provide a short review how to discuss the unitarity making use of
the BRST charge. Afterwards we make a detailed study of the most
general version of our action, and we come to the conclusion that
the model is not unitary, as we are unable to remove all the
negative norm states from the physical spectrum in a consistent way.
\end{abstract}
%\tableofcontents
\sect{Introduction} In the two previous papers
\cite{Capri:2005dy,Capri:2006ne}, the following action was
constructed
\begin{eqnarray}
  S_{phys} &=& S_{cl} +S_{gf}\;,\label{completeaction}\\
  S_{cl}&=&\int d^4x\left[-\frac{1}{4}F_{\mu \nu }^{a}F_{\mu \nu }^{a}+\frac{im}{4}(B-\overline{B})_{\mu\nu}^aF_{\mu\nu}^a
  +\frac{1}{4}\left( \overline{B}_{\mu \nu
}^{a}D_{\sigma }^{ab}D_{\sigma }^{bc}B_{\mu \nu
}^{c}-\overline{G}_{\mu \nu }^{a}D_{\sigma }^{ab}D_{\sigma
}^{bc}G_{\mu \nu
}^{c}\right)\right.\nonumber\\
&-&\left.\frac{3}{8}%
m^{2}\lambda _{1}\left( \overline{B}_{\mu \nu }^{a}B_{\mu \nu
}^{a}-\overline{G}_{\mu \nu }^{a}G_{\mu \nu }^{a}\right)
+m^{2}\frac{\lambda _{3}}{32}\left( \overline{B}_{\mu \nu
}^{a}-B_{\mu \nu }^{a}\right) ^{2}\right.\nonumber\\&+&\left.
\frac{\lambda^{abcd}}{16}\left( \overline{B}_{\mu\nu}^{a}B_{\mu\nu}^{b}-\overline{G}_{\mu\nu}^{a}G_{\mu\nu}^{b}%
\right)\left( \overline{B}_{\rho\sigma}^{c}B_{\rho\sigma}^{d}-\overline{G}_{\rho\sigma}^{c}G_{\rho\sigma}^{d}%
\right) \right]\;, \label{completeactionb}\\
S_{gf}&=&\int d^{4}x\;\left( \frac{\xi }{2}b^{a}b^{a}+b^{a}%
\partial _{\mu }A_{\mu }^{a}+\overline{c}^{a}\partial _{\mu }D_{\mu
}^{ab}c^{b}\right)\;.\label{lcg}
\end{eqnarray}
The bosonic fields $B_{\mu\nu}^a$, its conjugate
$\overline{B}_{\mu\nu}^a$ and the fermionic (ghost) fields
$G_{\mu\nu}^a$ and $\overline{G}_{\mu\nu}^a$ are antisymmetric in
their Lorentz indices and belong to the adjoint representation.
$\lambda^{abcd}$ is a gauge invariant quartic tensor coupling,
subject to a generalized Jacobi identity \cite{vanRitbergen:1998pn}
\begin{equation}\label{jacobigen}
    f^{man}\lambda^{mbcd}+f^{mbn}\lambda^{amcd}+f^{mcn}\lambda^{abmd}+f^{mdn}\lambda^{abcm}=0\,,
\end{equation}
and to the symmetry constraints
\begin{eqnarray}
\lambda^{abcd}=\lambda^{cdab} \;, \nonumber \\
\lambda^{abcd}=\lambda^{bacd} \;, \label{abcd}
\end{eqnarray}
while $\lambda_1$ and $\lambda_3$ are mass couplings\footnote{In
comparison with \cite{Capri:2005dy,Capri:2006ne}, we changed the
sign of $m$, $\lambda_3$, $\lambda_1$ and $\lambda^{abcd}$ to avoid
a number of minus signs.}.

\noindent To avoid confusion, let us mention here that we shall work
in Minkowski space throughout this paper, since we plan to come to
the canonical quantization. In \cite{Capri:2005dy,Capri:2006ne}, the
action was treated in Euclidean space.

\noindent The classical part of the action, $S_{cl}$, enjoys a
non-Abelian gauge invariance generated by
\begin{eqnarray}
\delta A_{\mu }^{a} &=&-D_{\mu }^{ab}\omega ^{b}\;,  \nonumber \\
\delta B_{\mu \nu }^{a} &=&gf^{abc}\omega ^{b}B_{\mu \nu }^{c}\;,
\nonumber
\\
\delta \overline{B}_{\mu \nu }^{a} &=&gf^{abc}\omega
^{b}\overline{B}_{\mu \nu }^{c}\;,
\nonumber \\
\delta G_{\mu \nu }^{a} &=&gf^{abc}\omega ^{b}G_{\mu \nu }^{c}\;,
\nonumber
\\
\delta \overline{G}_{\mu \nu }^{a} &=&gf^{abc}\omega
^{b}\overline{G}_{\mu \nu }^{c}\;, \label{gtm}
\end{eqnarray}
with $\omega^a$ parametrizing an arbitrary infinitesimal $SU(N)$
gauge transformation.

\noindent Quite obviously, the gauge model (\ref{completeaction})
did not come out of thin air. Our original motivation was based on
the quest for a dynamical mass generation mechanism in gauge
theories. We do not plan to give a complete overview of this issue,
but let us mention that this has been a research topic since long,
see e.g. \cite{Cornwall:1981zr} for a seminal work on this.

\noindent More recently, work appeared in which a dynamical gluon
mass was introduced phenomenologically based on the QCD sum rules
\cite{Chetyrkin:1998yr}. Such a mass can account for $\frac{1}{Q^2}$
power corrections in certain physical correlators
\cite{Chetyrkin:1998yr,Gubarev:2000eu,Gubarev:2000nz}. A natural
question arising is where this mass scale would originate from? The
authors of \cite{Gubarev:2000eu,Gubarev:2000nz} invoked the
condensation of the operator
\begin{equation}\label{m1}
    A^2_{\min}=(VT)^{-1}\min_{U\in SU(N)}\int d^4x
    \left(A_\mu^U\right)^2\,,
\end{equation}
since it is gauge invariant due to the minimization along the gauge
orbit\footnote{One should however be aware of the problem of gauge
(Gribov) ambiguities \cite{Gribov:1977wm,Semenov} for determining
the global minimum.}. As it is well known, a \emph{local} gauge
invariant dimension two operator does not exist in Yang-Mills gauge
theories. The nonlocality of (\ref{m1}) is best seen when it is
expressed as a series in Euclidean space \cite{Lavelle:1995ty}
\begin{eqnarray}\label{m2}
&&A_{\min }^{2} =\frac{1}{2VT}\int d^{4}x\left[ A_{\mu }^{a}\left( \delta _{\mu \nu }-\frac{%
\partial _{\mu }\partial _{\nu }}{\partial ^{2}}\right) A_{\nu
}^{a}-gf^{abc}\left( \frac{\partial _{\nu }}{\partial ^{2}}\partial
A^{a}\right) \left( \frac{1}{\partial ^{2}}\partial {A}^{b}\right)
A_{\nu }^{c}\right] +\ldots
\end{eqnarray}
which contains the inverse Laplacian $\frac{1}{\p^2}$ several times.
This is a nonlocal operator, as it can be immediately inferred from
its formal expression in $d$ dimensions through
\begin{equation}\label{invlap}
    \frac{1}{\p_x^2}f(x)=-\frac{\Gamma\left(\frac{d}{2}\right)}{2\pi^{\frac{d}{2}}(d-2)}\int
d^d y\frac{f(y)}{\vert x-y\vert^{d-2}}\,.
\end{equation}
\noindent All efforts so far were concentrated on the Landau gauge
$\p_\mu A_\mu=0$. The preference for this particular gauge is
obvious since the nonlocal expression (\ref{m2}) reduces to an
(integrated) local operator, more precisely
\begin{equation}\label{m3}
    \p_\mu A_\mu=0\Rightarrow A^2_{\min}=(VT)^{-1}\int d^4xA_\mu^2\,.
\end{equation}
In the case of a local operator like $A_\mu^2$, the Operator Product
Expansion (OPE), viz. short distance expansion, becomes applicable,
and consequently a measurement of the soft (infrared) part $\langle
A_\mu^2\rangle_{\mathrm{OPE}}$ becomes possible. Such an approach
was followed in e.g. \cite{Boucaud:2001st} by analyzing the
appearance of $\frac{1}{Q^2}$ power corrections in (gauge variant)
quantities like the gluon propagator or the strong coupling
constant, defined in a particular way, from lattice simulations. Let
us mention that already two decades ago attention was paid to
$\langle A_\mu^2\rangle_{\mathrm{OPE}}$ when the OPE was applied to
the propagators \cite{Lavelle:1988eg}. This condensate $\langle
A_\mu^2\rangle_{\mathrm{OPE}}$ can also be related to an effective
gluon mass, see e.g. \cite{Kondo:2001nq}.

\noindent A more direct approach to a determination of $\langle
A_\mu^2\rangle$ in the Landau gauge was presented in
\cite{Verschelde:2001ia,Dudal:2003vv}. In \cite{Verschelde:2001ia},
a meaningful effective potential for the condensation of the local
composite operator $A_\mu^2$ was constructed, giving evidence of
$\langle A_\mu^2\rangle\neq0$, and as a consequence a nonvanishing
gluon mass of a few hundred $MeV$ was found. The renormalizability
of this technique was proven to all orders of perturbation theory in
\cite{Dudal:2002pq}.

\noindent Effective gluon masses have found application in
phenomonological studies like
\cite{Parisi:1980jy,Halzen:1992vd,Field:2001iu}. Also lattice
simulations of the gluon propagator revealed the need for massive
parameters, when the obtained form factors are fitted by means of
functional forms
\cite{Marenzoni:1994ap,Langfeld:2001cz,Amemiya:1998jz,Bornyakov:2003ee}.
Other approaches to dynamical gluon masses are e.g.
\cite{Aguilar:2004sw, Aguilar:2006gr}. The estimates of these mass
parameters are grosso modo all in the same ballpark, ranging from a
few hundred $MeV$ up to $1.2\;GeV$.

\noindent It is perhaps important to spend a few words at clearing
up a common misconception. The concept of a dynamically generated
effective gluon mass does not necessarily entail that we are
considering massive gauge bosons that are belonging to the physical
spectrum, i.e. that are observable particles. At low energies,
perturbative QCD expressed in terms of gluons and quarks completely
fails, and the effective degrees of freedom become the hadrons. The
phenomena we are interested in, in casu the study of the condensates
and ensuing dynamical mass generation, occur in a energy window
located in between perturbative QCD and the confined region.
Perturbation theory still has its validity there, but it gets
corrected by nonperturbative effects like condensates. Due to the
lack of an explicit knowledge of the correct physical degrees of
freedom (the hadrons), we continue to use the gluons as effective
degrees of freedom, although we are already out of the energy regime
where these might be considered as asymptotic observables. If we
cross from high to low energies, the originally massless and
physical gluons will not stante pede become confined at the
confinement scale, but rather they will behave as a kind of massive
quasi particles before getting confined, and this happens at scales
that are phenomenologically relevant. This also means that unitarity
in terms of the gluons is not required or even desired. One expects
that quasi particles do have a finite lifetime and cannot be
observed as asymptotically free particles.

\noindent We have already explained the preferred role of the Landau
gauge, since in that case a gauge invariant meaning can be assigned
to $\langle A_\mu^2\rangle$. Obviously, since we are working in a
gauge theory, the condensates influencing physical quantities should
be at least gauge invariant. Therefore, it would be nice to have a
dimension 2 condensate that could also be treated in other gauges.
As the operator $A^2_{\min}$ remains nonlocal, it falls beyond the
applicability of the OPE. It is also unclear how e.g.
renormalizability or an effective potential approach could be
established for nonlocal operators. In most covariant gauges, we and
collaborators have discussed that other dimension two,
renormalizable and local operators exist. We showed that these
operators condense and give rise to a dynamical gluon mass, see
Table 1 and
\cite{Dudal:2003gu,Dudal:2003pe,Dudal:2003np,Browne:2003uv,Dudal:2003by,Dudal:2004rx,Browne:2004mk,Gracey:2004bk}.
Quite recently, it has also been shown that a class of nonlinear
covariant gauges enjoys the fact that $A_\mu^2$ is multiplicatively
renormalizable \cite{Lemes:2006aw}.
\begin{table}
\begin{center}
\begin{tabular}{|c|c|}
  \hline
  Gauge&Operator\\
  \hline\hline
  linear covariant &   $\frac{1}{2}A_\mu^a A_\mu^a$\\
  Curci-Ferrari & $\frac{1}{2}A_\mu^a
  A_\mu^a+\alpha\overline{c}^ac^a$\\
  maximal Abelian & $\frac{1}{2}A_\mu^{\beta} A_\mu^{\beta}+\alpha\overline{c}^{\beta}c^{\beta}$ \\
nonlinear class& $\frac{1}{2}A_\mu^a A_\mu^a$\\
  \hline
\end{tabular}
\caption{Gauges and their renormalizable dimension two operator}
\end{center}
\end{table}
In the maximal Abelian gauge, it was found that only the
off-diagonal gluons $A_\mu^\beta$ acquire a dynamical mass
\cite{Dudal:2004rx}, a fact qualitatively consistent with the
lattice results from \cite{Amemiya:1998jz,Bornyakov:2003ee}.  Let us
also mention that we have been able to make some connection between
the various gauges and their dimension two operators by constructing
renormalizable interpolating gauges and operators
\cite{Dudal:2004rx,Dudal:2005zr}. These can be used to obtain a
formal result on the gauge parameter independence of the
nonperturbative vacuum energy due to the condensation, which is
lower than the perturbative (zero) vacuum energy
\cite{Dudal:2003by}.

\noindent A certain disadvantage of the research so far is the
explicit gauge dependence of the used operator. We started looking
for a gauge invariant dimension two operator, which a fortiori needs
to be nonlocal. We would like to develop a consistent
(calculational) framework, hence we are almost forced to look for an
operator that can be localized by introducing a suitable set of
extra fields. From this perspective, $A^2_{\min}$ seems to be rather
inadequate as it is a infinite series of nonlocal terms. A perhaps
more appealing operator is \cite{Capri:2005dy}
\begin{equation}\label{m10}
    \mathcal{O}=\frac{1}{VT}\int d^{4}xF_{\mu \nu }^{a}\left[
\left( D^{2}\right) ^{-1}\right] ^{ab}F_{\mu \nu }^{b}\,.
\end{equation}
This operator found already use in the study of a dynamical mass
generation in 3-dimensional gauge theories \cite{Jackiw:1995nf}.
When we add the operator $\mathcal{O}$ to the Yang-Mills action via
\begin{equation}
S_{YM}-\frac{m^{2}}{4}\int d^{4}xF_{\mu \nu }^{a}\left[ \left(
D^{2}\right) ^{-1}\right] ^{ab}F_{\mu \nu }^{b}\,,
\end{equation}
we can localize it to
\begin{eqnarray}\label{m11}
&&S_{YM}+\int d^{4}x\left[\frac{im}{4}\left( B-\overline{B%
}\right) _{\mu \nu }^{a}F_{\mu \nu }^{a}+\frac{1}{4}\left(
\overline{B}_{\mu \nu }^{a}D_{\sigma }^{ab}D_{\sigma }^{bc}B_{\mu
\nu }^{c}-\overline{G}_{\mu \nu }^{a}D_{\sigma }^{ab}D_{\sigma
}^{bc}G_{\mu \nu }^{c}\right)\right]\,,
\end{eqnarray}
at the cost of introducing a set of extra fields
\cite{Capri:2005dy}.

\noindent The action (\ref{m11}) as it stands is however not
renormalizable, but we and collaborators have shown that the
generalized version (\ref{completeaction}) is renormalizable to all
orders in the class of linear covariant gauges, implemented through
$S_{gf}$, in \cite{Capri:2005dy,Capri:2006ne}. We have also
calculated several renormalization group functions to two loop
order, confirming the renormalizability at the practical level.
Various consistency checks were at our disposal in order to
establish the reliability of these results, e.g. the gauge parameter
independence of the anomalous dimension of gauge invariant
quantities like $g^2$, $\lambda^{abcd}$ or $m$, or the equality of
others, in accordance with the output of the Ward identities in
\cite{Capri:2005dy}. We refer the reader to
\cite{Capri:2005dy,Capri:2006ne} for all details concerning the
localization procedure or renormalizability analysis, as well as the
need for the extra couplings.

\noindent Furthermore, we have proven in \cite{Capri:2006ne} the
perturbative equivalence of the model (\ref{completeaction}) with
ordinary Yang-Mills theory in the case that $m\equiv0$. We notice
that this is a nontrivial statement due to the presence of the
quartic interaction $\sim \lambda^{abcd}$ in the extra fields. It
has an interesting corollary: because we employ a massless
renormalization scheme, in casu $\MSbar$, we can set the mass $m$
equal to zero to determine the renormalization group functions of
e.g. the coupling constant $g^2$, the gauge parameter $\xi$ or
original Yang-Mills fields. Since both theories are perturbatively
equivalent for $m\equiv0$, the already mentioned renormalization
group functions must be identical. This has indeed been confirmed by
the explicit results of \cite{Capri:2005dy,Capri:2006ne}. In
particular, our model is thus asymptotically free at high energies,
with or without a mass. At lower energies, nonperturbative effects
can set in, completely analogous to the Yang-Mills case.

\noindent Summarizing, we have thus found a classically gauge
invariant action, which at the quantum level can be renormalized to
all orders in at least the class of linear covariant gauges, and as
a bonus it is perturbatively equivalent with ordinary Yang-Mills
gauge theories for vanishing mass. We can now ask ourselves two
questions:
\begin{enumerate}
\item If we treat the mass $m$ as a given classical input, can we consider
our model as a candidate for a gauge theory with massive
excitations? Therefore, we should prove that the theory is unitary,
containing massive particles in a suitably defined asymptotic
physical subspace. The particles correspond to the elementary
excitations of the original fields. As it is well known, proving the
unitarity of gauge theories is not a trivial job. A well known proof
in the case of Yang-Mills theories based on the BRST symmetry
\cite{Becchi:1975nq,Tyutin}, is given in \cite{Kugo:1979gm}.

\item If we do not want to treat our model as one with a given
classical mass $m$, can we dynamically generate it in a
selfconsistent way in this case? Said otherwise, can we develop a
method to find a reasonable gap equation for this mass? At high
energies, the model is massless and the same as Yang-Mills theory,
but it might develop a dynamical mass scale at lower energies,
without spoiling the gauge invariance. We cannot add mass terms to
the Yang-Mills action without spoiling the gauge invariance or
renormalizability, but we can add mass terms to our model. Just as
for Yang-Mills theories, we expect our model to be confining at
lower energies. As we have already mentioned, the original fields
can develop a behaviour different from the one expected from
perturbation theory in the energy regime in between confinement and
the perturbatively accessible high energy region. For example, a
gauge invariant mass parameter could be dynamically generated,
thereby modifying the propagators in a nonperturbative fashion, and
this without the need that these describe asymptotically observable
physical particles.
\end{enumerate}
In this paper, we shall provide an answer to the first question by
studying the massive gauge model (\ref{completeaction}). More
precisely, we shall quantize the model canonically to have a clear
particle interpretation of the quantum fields, and we shall find out
whether it is possible to define a physical subspace of states
endowed with a positive norm. Naively, one might expect the model to
be unitary, because the action (\ref{completeaction}) enjoys a BRST
symmetry, generated by the nilpotent operator $s$,
\begin{eqnarray}
s A_{\mu }^{a} &=&-D_{\mu }^{ab}c ^{b}\;,  \nonumber \\
s c^{a} &=&\frac{g}{2}f^{abc}c^ac ^{b}\;,  \nonumber \\
 s B_{\mu \nu }^{a} &=&gf^{abc}c ^{b}B_{\mu \nu }^{c}\;,
\nonumber
\\
s \overline{B}_{\mu \nu }^{a} &=&gf^{abc}c ^{b}\overline{B}_{\mu \nu
}^{c}\;,
\nonumber \\
s G_{\mu \nu }^{a} &=&gf^{abc}c ^{b}G_{\mu \nu }^{c}\;, \nonumber
\\
s \overline{G}_{\mu \nu }^{a} &=&gf^{abc}c ^{b}\overline{G}_{\mu
\nu }^{c}\;,\nonumber\\
s\overline{c}^{a} &=&b^a\;,  \nonumber \\
s b^{a} &=&0\,,
 \label{brst3}\nonumber\\
 s^2&=&0\,.
\end{eqnarray}
It should not come as a surprise that we shall heavily rely on this
BRST symmetry to discuss the unitarity of the model. The paper is
organized as follows. In section 2, we review how a sensible
physical subspace can be defined by using the \emph{free} BRST
charge \cite{Slavnov:1989jh,Frolov:1989az}. As a warming up
exercise, we apply the results of section 2 to the well known
canonical quantization of Yang-Mills gauge theories in section 3,
before turning to the explicit quantization of the gauge model
(\ref{completeaction}) in section 4. In section 5, we discuss the
presence of some extra symmetries which allow to reduce the physical
subspace further. Section 6 is devoted to the (free) classical
equations of motion and the Fourier decomposition of the solutions.
We shall encounter the problem of ``multipoles'', since the free
equations of motion couple different fields to each other. This
gives rise to higher derivative decoupled equations of motion. We
also pay attention to the BRST charge and Hamiltonian. In section 7,
we discuss how to derive the commutation relations between the
creation and annihilation operators, without using the brackets
between the fields and their conjugate momenta, which we want to
avoid, since not all the Fourier components of the fields and
momenta are independent. Once this is done, we come to the
conclusion that the massive gauge model (\ref{completeaction}) is
not unitary, as we end up with negative norm modes in the physical
subspace. We are unaware of any step to further reduce this subspace
\emph{in a consistent way}\footnote{That means compatible with the
interactions of the model.} to remove these unwanted modes. We end
with some conclusions in section 8.

\sect{A constructive approach to the question of unitarity in gauge
theories} In this section, we shall review how we can construct the
action $S$ for an interacting gauge theory, if we have a free theory
$S_0$ at our disposal, together with a nilpotent symmetry generator
$s_0$, so that $s_0S_0=0$. The content of this section is mainly
based on \cite{Slavnov:1989jh,Frolov:1989az}, although here and
there we adapted the proofs. We shall only be concerned with the
non-reducible case in this paper.

\noindent Let us thus start from the free action $S_0$. This action
contains a set of fields, appearing quadratically. It is given that
$S_0$ enjoys a BRST symmetry $s_0$, with corresponding nilpotent
charge $\mathcal{Q}_0$. As a standard example, we can consider the
free part of a gauge theory in a particular gauge, with its
corresponding gauge fixing part. For example, in the linear
covariant gauge we have
\begin{eqnarray}\label{ym1}
  S_0&=&\int d^4x\left[-\frac{1}{4}\left(\p_\mu A_\nu^a-\p_\nu A_\mu^a\right)^2+b^{a}
\partial _{\mu }A^{\mu a}+\overline{c}^{a}\p^2 c^{a}+\frac{\xi}{2}
b^ab^a\right]\,.
\end{eqnarray}
The free BRST symmetry is generated by
\begin{eqnarray}\label{ym2}
  s_0A_\mu^a &=& -\p_\mu c^a\,,\nonumber \\
  s_0 c^a &=&0\,, \nonumber \\
  s_0 \occ^a &=& b^a\,, \nonumber \\
  s_0 b^a &=& 0\,,\nonumber\\
    s_0^2&=&0\,.
\end{eqnarray}
We mention that also the free ``ghost part'' has to be included in
$S_0$. We may define the ghost charge $\mathcal{Q}_{gh}$. In
\cite{Slavnov:1989jh,Frolov:1989az}, the ghost charge is not used.
We may use it anyhow in the definition of the physical subspace.
However, this requirement is a bit redundant. A BRST cohomological
analysis (see later) will eventually learn that a physical state
counts neither ghosts nor anti-ghosts in the case of Yang-Mills
gauge theories.

\noindent Unitarity means that we start from a physical state space
$\mathcal{H}_{\mbox{\tiny phys}}$, which is a subspace of the total
Hilbert state space $\mathcal{H}$. $\mathcal{H}_{\mbox{\tiny phys}}$
should of course be endowed with a positive norm in order to have a
sensible probabilistic interpretation of the quantum theory. If we
let the states of $\mathcal{H}_{\mbox{\tiny phys}}$ interact, we
must end up again in the (same) subspace $\mathcal{H}_{\mbox{\tiny
phys}}$. Nonphysical states, which can have negative norms, may
contribute to the $\mathcal{S}$-matrix in internal processes, but
they cannot appear in the observable sector (the ``out''-space),
unless perhaps in zero norm combinations.

\noindent Consequently, two questions need to be answered:
\begin{enumerate}
\item How do we define the physical subspace $\mathcal{H}_{\mbox{\tiny phys}}$?
\item Do the states in the physical subspace $\mathcal{H}_{\mbox{\tiny phys}}$ possess a positive norm?
\end{enumerate}
Let us first explain how we define our physical subspace, starting
from the free action. A state $\vert\psi_p\rangle$ is called
physical if
\begin{equation}\label{1}
    \ph\in\mathcal{H}_{\mbox{\tiny phys}}\Leftrightarrow\mathcal{Q}_0\ph=0\;,\quad\ph\neq
\vert\ldots\rangle+\mathcal{Q}_0\vert\ldots\rangle\,.
\end{equation}
Physical states are thus defined from the ``free'' BRST charge $
\mathcal{Q}_0$. Since $\mathcal{Q}_0$ is supposed to be nilpotent,
states of the form $\mathcal{Q}_0\vert\ldots\rangle$ are trivially
annihilated by $\mathcal{Q}_0$. We notice that these have zero
norm\footnote{The BRST charge can be chosen to be Hermitian.}. We
can identify them with the trivial state, more mathematically
speaking this amounts to consider the $\mathcal{Q}_0$ cohomology. In
the usual terminology, we define $\mathcal{Q}_0$-closed and
$\mathcal{Q}_0$-exact states by
\begin{eqnarray}
  \ph \mbox{ is $\mathcal{Q}_0$-closed} &\Leftrightarrow&  \mathcal{Q}_0\ph=0\Leftrightarrow \ph\in\mbox{Ker}\mathcal{Q}_0\,,\\
   \ph \mbox{ is $\mathcal{Q}_0$-exact} &\Leftrightarrow&  \ph=\mathcal{Q}_0\vert\phi\rangle\Leftrightarrow
\ph\in\mbox{Im}\mathcal{Q}_0\,.
\end{eqnarray}
Since $\mathcal{Q}_0^2=0$, every exact state is trivially closed,
meaning that $\mbox{Im}\mathcal{Q}_0\subset\mbox{Ker}\mathcal{Q}_0$.
Hence, we can reexpress the condition (\ref{1}) as
\begin{eqnarray}\label{phys}
  \ph\in\mathcal{H}_{\mbox{\tiny phys}} &\Leftrightarrow&  \ph\in
\mbox{cohom}\mathcal{Q}_0\equiv\frac{\mbox{Ker}\mathcal{Q}_0}{\mbox{Im}\mathcal{Q}_0}\,.
\end{eqnarray}
For the moment, we leave open the (key) question whether these
states $\ph$ have a positive norm.

\noindent The next problem is whether we can construct an action $S$
compatible with unitarity? Starting from the free action $S_0$, we
can complement it order by order with terms in the coupling
constant(s), so that
\begin{equation}\label{2}
    S=S_0+S_1+S_2+\ldots\,.
\end{equation}
The question becomes how to determine the interaction terms $S_1$,
$S_2$, ..., such that $S$ describes a unitary model? More precisely,
having defined a physical subspace by means of (\ref{phys}), we
would like to construct the action $S$ such that the subspace
defined by (\ref{phys}) is maintained under time evolution. In the
operator language, we must therefore require that the time evolution
operator $\mathcal{S}$, given by
\begin{equation}
\mathcal{S}=\mathcal{T}\left[e^{-i\int_{-\infty}^{+\infty}H_{int}(t)dt}\right]\,,
\end{equation}
with $\mathcal{T}$ the usual time-ordering operation, commutes with
the operator $\mathcal{Q}_0$. Then clearly the $\mathcal{S}$-matrix
will be unitary, as states evolved w.r.t. $\mathcal{S}$ will
asymptotically again belong to the (same) physical subspace
$\mathcal{H}_{\mbox{\tiny phys}}$.

\noindent In order to solve the previous requirement, we prefer to
work in the path integral language rather than in the operator
language. Let us thus rephrase the previous requirement in the path
integral language. From the LSZ reduction formulae, see
\cite{Peskin:1995ev} and \cite{Lehmann:1954rq} for the original
paper, we know that the $\mathcal{S}$-matrix elements\footnote{We
can restrict ourselves to the connected $\mathcal{S}$-matrix
elements.} are determined by the (connected) amputated $n$-point
Green functions, put on-shell. In a rough notation, we can write
\begin{equation}\label{mat1}
    \langle\vk_1^\prime,\ldots,\vk_m^\prime\vert\mathcal{S}\vert\vk_1,\ldots,\vk_n\rangle\sim
\langle\Theta\vert\mathcal{T}\left[\phi_m\phi_n\right]\vert\Theta\rangle\,,
\end{equation}
where $\sim$ symbolizes all the necessary prefactors, putting it
on-shell, amputating and Fourier transforming to momentum space.
$\phi_m$ and $\phi_n$ are certain functionals of the fields, leading
to a $(m+n)$-point function.

\noindent Starting from a generic physical state
$\vert\vk_1,\ldots,\vk_n\rangle$  with the property
\begin{equation}\label{mat1bis}
\mathcal{Q}_0 \vert\vk_1,\ldots,\vk_n\rangle=0\,,
\end{equation}
we are wondering which condition will assure that
\begin{equation}\label{mat2}
    \langle\vk_1^\prime,\ldots,\vk_m^\prime\vert\mathcal{Q}_0\mathcal{S}\vert\vk_1,\ldots,\vk_n\rangle
\stackrel{?}{=}\langle\vk_1^\prime,\ldots,\vk_m^\prime\vert\mathcal{S}\mathcal{Q}_0\vert\vk_1,\ldots,\vk_n\rangle=0\,,
\end{equation}
where the last equality follows from (\ref{mat1bis}). As it is well
known, we can express the $\mathcal{T}$-ordered product with the
path integral, so that we find\footnote{We shall use the notation
$\mathcal{Q}_0$ for the charge, eventually written in terms of
creation/annihilation operators, while $\mathcal{B}_0$ represents
the functional analog of $\mathcal{Q}_0$. }
\begin{eqnarray}\label{mat3}
    &&\langle\vk_1^\prime,\ldots,\vk_m^\prime\vert\mathcal{Q}_0\mathcal{S}\vert\vk_1,\ldots,\vk_n\rangle
\sim
\langle\Theta\vert\mathcal{T}\left[\left(\mathcal{Q}_0\phi_m\right)\phi_n\right]\vert\Theta\rangle=\int
dX \left(\mathcal{B}_0\phi_m\right)\phi_n e^{iS}\,,
\end{eqnarray}
where $X$ represents all the fields. Now, we can write
\begin{eqnarray}\label{mat4}
    \int dX \left(\mathcal{B}_0\phi_m\right)\phi_n e^{iS}&=&\int dX \mathcal{B}_0\left(\phi_m\phi_n\right)
e^{iS}\,,
\end{eqnarray}
since $\mathcal{B}_0\phi_n=0$ by virtue of (\ref{mat1bis}).

\noindent Consider the path integral
\begin{eqnarray}\label{mat5}
    \int dX \Phi e^{iS}\,,
\end{eqnarray}
for an arbitrary functional $\Phi$ of the fields. We perform the
transformation of the path integral variables
\begin{equation}\label{mat6}
    X=X^\prime+\epsilon \mathcal{B}_0X^\prime\,,\qquad\qquad\epsilon
\mbox{ infinitesimal}\,.
\end{equation}
As $\mathcal{B}_0$ induces a linear transformation, there is no
associated Jacobian, and we find
\begin{eqnarray}\label{mat7}
    \int dX \Phi e^{iS}=\int dX^\prime \Phi^\prime e^{iS^\prime}+\epsilon\int dX^\prime \mathcal{B}_0\left(\Phi^\prime
e^{iS^\prime}\right)\,.
\end{eqnarray}
Dropping the prime again, we find
\begin{eqnarray}\label{mat8}
    \int dX \mathcal{B}_0\left(\Phi
e^{iS}\right)=0\,.
\end{eqnarray}
Taking a look at (\ref{mat8}), we can be certain that (\ref{mat4})
holds when we impose
\begin{equation}\label{mat9}
   \mathcal{B}_0e^{iS}=0\,.
\end{equation}

\noindent In order to proceed, we notice that it is in principle
sufficient that (\ref{mat9}) is fulfilled on-shell as the
$\mathcal{S}$-matrix is of course considered on-shell. At the level
of the action however, we must require that it holds off-shell. Let
us introduce a (very) condensed notation for the action of
$\mathcal{B}_0$
\begin{equation}\label{6}
    \mathcal{B}_0=\Delta\phi_0\frac{\p}{\p\phi}\,.
\end{equation}
Implementing (\ref{mat9}) at lowest order and making it valid
off-shell means that
\begin{eqnarray}
    \label{7b}\mathcal{B}_0S_1&=&0\underbrace{\Rightarrow}_{\mathrm{off-shell}}\Delta\phi_0\frac{\p}{\p\phi}S_1=-\Delta\phi_1\frac{\p
S_0}{\p
    \phi}\,.
\end{eqnarray}
We already see here that $\mathcal{B}_0$ will get adapted, more
precisely we can introduce the modified operator $\mathcal{B}$ by
\begin{equation}\label{8}
    \mathcal{B}_0\equiv\Delta\phi_0\frac{\p}{\p\phi}\rightarrow
\mathcal{B}=\mathcal{B}_0+\Delta\phi_1\frac{\p}{\p\phi}\,,
\end{equation}
so that
\begin{equation}\label{9}
    \mathcal{B}(S_0+S_1)=0\,.
\end{equation}
We remind here that all $\mathcal{Q}$'s ($\mathcal{B}$'s) are
Grassmann operators.

\noindent Since $\mathcal{B}_0$ is nilpotent, we can act with it on
(\ref{7b}) to find that
\begin{eqnarray}\label{11}
  0 &=& \mathcal{B}_0^2S_1=\mathcal{B}_0\left(-\Delta\phi_1\frac{\p}{\p\phi}S_0\right) \nonumber\\
  &=&-\Delta\phi_0\frac{\p}{\p \phi}(\Delta\phi_1)\frac{\p}{\p \phi}S_0+\Delta\phi_1\Delta\phi_0\frac{\p^2}{\p
  \phi^2}S_0\,.
\end{eqnarray}
Acting with $\frac{\p}{\p\phi}$ on
\begin{equation}
    0=\mathcal{B}_0S_0=\Delta\phi_0\frac{\p}{\p\phi}S_0\,,
\end{equation}
yields
\begin{equation}\label{10}
    \frac{\p}{\p\phi}(\Delta\phi_0)\frac{\p}{\p\phi}S_0+\Delta\phi_0\frac{\p^2}{\p\phi^2}S_0=0\,.
\end{equation}
Combination of (\ref{11}) and (\ref{10}) learns
\begin{equation}\label{12}
    \Delta\phi_0\frac{\p}{\p\phi}(\Delta\phi_1)\frac{\p}{\p\phi}S_0+\Delta\phi_1\frac{\p}{\p\phi}(\Delta\phi_0)\frac{\p}{\p\phi}S_0=0\,,
\end{equation}
from which we infer that
\begin{equation}\label{13}
    \Delta\phi_0\frac{\p}{\p\phi}(\Delta\phi_1)\frac{\p}{\p\phi}+\Delta\phi_1\frac{\p}{\p\phi}(\Delta\phi_0)\frac{\p}{\p\phi}=0\,.
\end{equation}
The identity (\ref{13}) expresses nothing more than the nilpotency
of $Q=Q_0+Q_1$, given by (\ref{8}) since, at lowest order
\begin{eqnarray}\label{14}
    \mathcal{B}^2&=&(\Delta\phi_0\frac{\p}{\p\phi}+\Delta\phi_1\frac{\p}{\p\phi})(\Delta\phi_0\frac{\p}{\p\phi}+\Delta\phi_1\frac{\p}{\p\phi})\nonumber\\
    &=&\Delta\phi_1\frac{\p}{\p\phi}(\Delta\phi_0)\frac{\p}{\p\phi}+\Delta\phi_1\Delta\phi_0\frac{\p^2}{\p\phi^2}+\Delta\phi_0\Delta\phi_1\frac{\p^2}{\p\phi^2}+\Delta\phi_0\frac{\p}{\p\phi}(\Delta\phi_1)\frac{\p}{\p\phi}+HOT
\nonumber\\
&=&\Delta\phi_1\frac{\p}{\p\phi}(\Delta\phi_0)\frac{\p}{\p\phi}+\Delta\phi_0\frac{\p}{\p\phi}(\Delta\phi_1)\frac{\p}{\p\phi}=(\ref{13})=0\,,
\end{eqnarray}
where we used for example the nilpotency of $\mathcal{B}_0$. We
dropped the last term as it is of higher order.

\noindent We notice that the potential solution of (\ref{7b}) is
apparently restrained by the condition that it is invariant under a
nilpotent operator $Q$ ($\mathcal{B}$), which reduces to $Q_0$
($\mathcal{B}_0$) in the free limit.

\noindent This construction can be continued at higher order. One
proves that the action at order $n$,
\begin{equation}\label{14bis}
    S=S_0+S_1+\ldots + S_n\,,
\end{equation}
is the solution of
\begin{equation}\label{15}
    \Delta\phi_0\frac{\p}{\p\phi}S_n+\Delta\phi_1\frac{\p}{\p\phi}S_{n-1}+\ldots+\Delta\phi_n\frac{\p}{\p\phi}S_{0}=0\,,
\end{equation}
where consistency demands that the BRST operator,
\begin{equation}\label{16}
    \mathcal{B}=\mathcal{B}_0+\ldots
+\mathcal{B}_n=\Delta\phi_0\frac{\p}{\p\phi}+\ldots+\Delta\phi_n\frac{\p}{\p\phi}\,,
\end{equation}
is nilpotent at the considered order $n$, thus
\begin{equation}\label{17}
    \mathcal{B}^2=0\;\;\;\;(\textrm{or } Q^2=0)\,.
\end{equation}
By construction, the final action (\ref{14bis}) shall be invariant
under the BRST symmetry generated by (\ref{16}).

\noindent To make things a bit more comprehensible, let us work out
the procedure at second order. We hence demand that (\ref{2}) is
consistent with (\ref{mat9}), and this extended to the off-shell
level, meaning that
\begin{equation}\label{a}
    i\mathcal{B}_0S_2-\frac{1}{2}\mathcal{B}_0(S_1S_1)=-i\widetilde{\Delta\phi_2}\frac{\p
S_0}{\p\phi}\,.
\end{equation}
The complex unity $i$ in the r.h.s. as well as the
$\widetilde{\;}$-notation are merely introduced for later
convenience. Using (\ref{7b}), we may rewrite (\ref{a}) as
\begin{equation}\label{a2}
    i\Delta\phi_0\frac{\p S_2}{\p\phi}+\Delta\phi_1\frac{\p
    S_0}{\p\phi}S_1=-i\widetilde{\Delta\phi_2}\frac{\p
S_0}{\p\phi}\,.
\end{equation}
Next, using Wick's theorem, we can write\footnote{This operation is
understood within the path integral.}
\begin{equation}\label{wick}
\Delta\phi_1\frac{\p
    S_0}{\p\phi}S_1=i\Delta\phi_1\frac{\p S_1}{\p\phi}+i\widetilde{\Delta\phi_1}\frac{\p
S_0}{\p\phi}\,,
\end{equation}
since roughly said, $\frac{\p S_0}{\p
\phi}\sim\mathcal{D}^{-1}\times\phi$, and a ``contraction'' of this
with a $\phi$ from $S_1$ will give rise to a
$i\mathcal{D}^{-1}\times\mathcal{D}$ with $\mathcal{D}$ a free
propagator. All other terms are taken together in $\widetilde{\Delta
\phi_1}$.

\noindent Upon taking (\ref{a2}) and (\ref{wick}) into account, we
come to the conclusion that
\begin{equation}\label{a3}
\Delta\phi_0\frac{\p S_2}{\p \phi}+\Delta\phi_1\frac{\p S_1}{\p\phi
}=-\Delta\phi_2\frac{\p S_0}{\p\phi}
\end{equation}
in order to have the condition (\ref{a}) fulfilled. We defined
\begin{equation}\label{a3}
    \Delta\phi_2=\widetilde{\Delta\phi_2}+\widetilde{\Delta\phi_1}\,.
\end{equation}
Analogously at is was proven in (9) to (14), the nilpotency of
$Q_0+Q_1$ leads to the nilpotency of $Q_0+Q_1+Q_2$ as a consistency
requirement.

\noindent Of course, there is no guarantee that the foregoing
``bottom top'' construction of the complete action will end at a
finite order. Given that it ends at a finite order, it could still
be a very cumbersome job to actually get the nilpotent BRST charge
and corresponding action. The situation becomes much more appealing
when we already have at our disposal a complete action, with a
nilpotent charge generating a symmetry. If the interaction is
switched off by setting all coupling constants equal to zero, we
obtain the free action, with a free nilpotent charge. When the above
``bottom top'' machinery is unleashed, the complete original action
and its BRST symmetry generator shall quite evidently be a solution
to the iterative procedure. From this viewpoint, we have a ``top
bottom'' approach to unitarity for actions with nilpotent BRST
charge, when they are ``reduced'' to their free counterpart.

\sect{Unitarity of Yang-Mills gauge theories using the BRST charge}
We should still provide an answer to question 2, namely do the
states that are annihilated by the free BRST charge have a positive
norm? It is well known that this is the case for Yang-Mills gauge
theories. For completeness, let us nevertheless repeat the argument.
This will allow for a comparison with Yang-Mills theories when we
start analyzing our generalized model.

\noindent We shall base ourselves on \cite{Henneaux:1992ig} for this
particular job\footnote{We shall however use other conventions than
those of \cite{Henneaux:1992ig}.}. We opt to work in the Feynman
gauge for simplicity ($\xi=1$ in (\ref{ym1})). Let us first
determine the conjugate momenta of all fields.
\begin{eqnarray}\label{br1}
  \pi_{i}^a &=& -F_{0i}^a\,,\nonumber \\
  \pi^{0,a} &=& b^a\,,\nonumber\\
\pi^a_c&=&\p_0\occ^a\,, \nonumber\\
\pi^a_{\occ}&=&-\p_0 c^a\,,
\end{eqnarray}
so that quantization requires
\begin{eqnarray}\label{br2}
    \left[A_i^a(\vec{x},t),\pi_{j}^b(\vec{y},t)\right]&=&i\delta^{ab}g_{ij}\delta^{(3)}(\vec{x}-\vec{y})\,,\nonumber\\
    \left[A_0^a(\vec{x},t),\pi^{0,b}(\vec{y},t)\right]&=&i\delta^{ab}\delta^{(3)}(\vec{x}-\vec{y})\,,\nonumber\\
    \Bigl\lbrace c^a(\vec{x},t),\pi^{b}_c(\vec{y},t)\Bigr\rbrace&=&i\delta^{ab}\delta^{(3)}(\vec{x}-\vec{y})\,,\nonumber\\
\Bigl\lbrace\occ^a(\vec{x},t),\pi^{b}_{\occ}(\vec{y},t)\Bigr\rbrace&=&i\delta^{ab}\delta^{(3)}(\vec{x}-\vec{y})\,,\nonumber\\
\textrm{other (anti-)commutators trivial}\,.
\end{eqnarray}
We mention that the classical equations of motion are
\begin{eqnarray}\label{br3}
    \p^2 A_\mu^a&=&0\,,\nonumber\\
\p^2c^a&=&\p^2\occ^a=0\,,\nonumber\\
b^a&=&-\p^\mu A_{\mu}^a\,,
\end{eqnarray}
and that we use the hermiticity assignment
\begin{eqnarray}\label{br4}
    c^\dagger&=&c\,,\qquad\qquad\occ^\dagger=-\oc\,.
\end{eqnarray}
We propose the following Fourier decompositions\footnote{We
suppressed the global color indices.}
\begin{eqnarray}\label{br5}
    A_0(x)&=&\int\frac{d^3k}{(2\pi)^3}\frac{1}{2\omega_k}\left(a_0(k)e^{-ikx}+a^\dagger_0(k)e^{ikx}\right)\,,\nonumber\\
A_i(x)&=&\int\frac{d^3k}{(2\pi)^3}\frac{1}{2\omega_k}\sum_{m=1}^3\left(a_m(k)\varepsilon_i^m(\vec{k})e^{-ikx}+a^\dagger_m(k)\varepsilon_i^m(\vec{k})e^{ikx}\right)\,,\nonumber\\
b(x)\equiv\pi_0(x)&=&i\int\frac{d^3k}{(2\pi)^3}\frac{1}{2}\left((a_0(k)-a_3(k))e^{-ikx}-(a_0^\dagger(k)-a_3^\dagger(k))e^{ikx}\right)\,,\nonumber\\
c(x)&=&\int
\frac{d^3k}{(2\pi)^3}\frac{1}{2\omega_k}\left(\eta(k)e^{-ikx}+\eta^\dagger(k)e^{ikx}\right)\,,\nonumber\\
\occ(x)&=&\int
\frac{d^3k}{(2\pi)^3}\frac{1}{2\omega_k}\left(\overline{\eta}(k)e^{-ikx}-\overline{\eta}^\dagger(k)e^{ikx}\right)\,.
\end{eqnarray}
The polarization vectors $\varepsilon_i^{(m)}(\vk)$ form an
orthonormal set,
$\varepsilon_i^{(m)}(\vk)\varepsilon_i^{(n)}(\vk)=\delta^{mn}$ with
$\varepsilon_i^{(3)}(\vk)=\frac{k_i}{|\vec{k}|}$. We shall assume
that the particles move along the $z$-axis, so that
$\varepsilon_i^j=\delta_{ij}$.

\noindent Implementing (\ref{br2}), we must require the following
(anti-)commutation rules
\begin{eqnarray}\label{br6}
\left[a_0(k),a_0(q)\right]&=&-2(2\pi)^3\omega_k\delta^{(3)}(\vec{k}-\vec{q})\,,\nonumber\\
\left[a_m(k),a_n(q)\right]&=&2(2\pi)^3\omega_k\delta^{(3)}(\vec{k}-\vec{q})\delta_{mn}\,,\nonumber\\
    \Bigl\lbrace
\overline{\eta}(k),\eta^\dagger(q)\Bigr\rbrace&=&-2(2\pi)^3\omega_k\delta^{(3)}(\vec{k}-\vec{q})\,,
\nonumber\\
\Bigl\lbrace\eta(k),\overline{\eta}^\dagger(q)\Bigr\rbrace&=&-2(2\pi)^3\omega_k\delta^{(3)}(\vec{k}-\vec{q})\;.
\end{eqnarray}
For later use, let us already introduce the operator
\cite{Henneaux:1992ig}
\begin{equation}\label{br7}
    \mathcal{N}=\int\frac{d^3k}{(2\pi)^3}\frac{1}{2\omega_k}\left(-a_0^\dagger(k)
a_0(k)+a_3^\dagger(k)a_3(k)-\eta^\dagger(k)\overline{\eta}(k)-\overline{\eta}^\dagger(k)\eta(k)\right)\,,
\end{equation}
which counts the unphysical modes.

\noindent We are now ready to express the BRST charge in terms of
the creation/annihilation operators. The BRST Noether current is
given by
\begin{eqnarray}
  \mathcal{J}_0^\mu &=& F^{\mu\nu,a}\p_\nu c^a-b^a\p_\mu c^a\,,
\end{eqnarray}
which leads to the charge
\begin{eqnarray}\label{br9}
    \mathcal Q_0&=&\int d^3x\left(c^a\p^0 b^a -b^a\p^0c^a\right)\,,
\end{eqnarray}
where use was made of the classical equation of motion
\begin{equation}\label{br10}
    \p_\mu F^{\mu\nu,a}=\p^\nu b^a\,.
\end{equation}
After substitution of (\ref{br5}) in (\ref{br9}), the BRST charge is
expressed as
\begin{eqnarray}\label{br11}
    \mathcal Q_0&=&\int \frac{d^3k}{(2\pi)^3}\left[
\left(a_0^\dagger(k)-a_3^\dagger(k)\right)\eta(k)+\eta^\dagger(k)\left(\vphantom{a_0^\dagger}a_0(k)-a_3(k)\right)\right]\,.
\end{eqnarray}
If we define
\begin{equation}\label{br12}
    \mathcal
R=\frac{1}{2}\int\frac{d^3k}{(2\pi)^3}\frac{1}{4\omega_k^2}\left[\left(a_0^\dagger(k)+a_3^\dagger(k)\right)\overline{\eta}(k)+\overline{\eta}^\dagger(k)\left(a_0(k)+a_3(k)\right)\right]\,,
\end{equation}
then a little algebra yields
\begin{equation}\label{br13}
    \mathcal{N}=\Bigl\lbrace\mathcal{Q}_0,\mathcal R\Bigr\rbrace\,.
\end{equation}
The fact that the ``nonphysical'' counting operator $\mathcal{N}$ is
BRST exact is a very powerful result \cite{Henneaux:1992ig}. Assume
that $\ph$ is constrained by
\begin{equation}\label{br14}
    \mathcal{Q}_0\ph=0\,,
\end{equation}
and that it contains $n\neq0$ unphysical modes, i.e.
\begin{equation}\label{br15}
    \mathcal{N}\ph=n\ph\,,
\end{equation}
then consequently
\begin{eqnarray}\label{br16}
    \ph &=&
\frac{\mathcal{N}}{n}\ph=\frac{1}{n}\left(\mathcal
Q_0\mathcal{R}+\mathcal{R}\mathcal Q_0\right)\ph=\mathcal
Q_0\left(\frac{1}{n}\mathcal{R}\ph\right)\,,
\end{eqnarray}
meaning that a state $\ph$ annihilated by the BRST charge and
containing nonphysical modes is a fortiori BRST exact, and hence it
is zero in the physical cohomology. Said otherwise, physical states
do not contain unphysical modes. The physical subspace
$\mathcal{H}_{\mbox{\tiny phys}}$ of Yang-Mills gauge theories does
only contain the 2 transverse polarizations of the gauge field,
whereas the scalar and longitudinal polarizations cancel with the
ghost degrees of freedom.

\noindent In \cite{Slavnov:1989jh}, a different proof was presented
of the fact that a state annihilated by a BRST charge
$\mathcal{Q}_0$ of the form (\ref{br11}) containing nonphysical
modes, must have zero norm. However, the cohomological approach used
in e.g. \cite{Henneaux:1992ig} is somewhat more elegant.

\sect{Application to the massive gauge model: preliminary remarks}
Setting the couplings $g$ and $\lambda^{abcd}$ equal to zero in
(\ref{completeaction}), we are considering the quadratic action
\begin{eqnarray}\label{last1}
  S_0&=&\int d^4x\left[-\frac{1}{4}\left(\p_\mu A_\nu^a-\p_\nu A_\mu^a\right)^2+\frac{im}{4}(B-\overline{B})_{\mu\nu}^a\left(\p^\mu A^{\nu a}-\p^\nu A^{\mu a}\right)
  \right.\nonumber\\&+&\left.\frac{1}{4}\left( \overline{B}_{\mu \nu
}^{a}\p^2 B^{\mu \nu, a}-\overline{G}_{\mu \nu }^{a}\p^2 G^{\mu \nu
a}\right)-\frac{3}{8}%
m^{2}\lambda _{1}\left( \overline{B}_{\mu \nu }^{a}B^{\mu \nu
a}-\overline{G}_{\mu \nu }^{a}G^{\mu \nu a}\right)
\right.\nonumber\\
&+&\left.m^{2}\frac{\lambda _{3}}{32}\left( \overline{B}_{\mu \nu
}^{a}-B_{\mu \nu }^{a}\right) ^{2}+b^{a}%
\partial _{\mu }A^{\mu a}+\overline{c}^{a}\p^2 c^{a}+\frac{\xi}{2} b^ab^a\right]
\end{eqnarray}
This action enjoys the free BRST symmetry generated by
\begin{eqnarray}\label{ourbrst}
  s_0A_\mu^a &=& -\p_\mu c^a\,,\nonumber \\
  s_0 c^a &=&0 \,,\nonumber \\
  s_0 B_{\mu\nu}^a &=&s_0 \overline{B}_{\mu\nu}^a=s_0 G_{\mu\nu}^a=s_0 \overline{G}_{\mu\nu}^a=0\,, \nonumber \\
  s_0 \occ^a &=& b^a \,,\nonumber \\
  s_0 b^a &=& 0\,,
\end{eqnarray}
where clearly
\begin{equation}\label{nilp}
    s_0^2=0\,.
\end{equation}
We can hence apply the results of section 2 to the action
(\ref{completeaction}). The only thing left to prove is that there
exist a physical subspace with positive norm. This subspace is
certainly annihilated by the (free) BRST charge, but nothing
prevents us from using other available symmetries to further reduce
the physical subspace. In the next section, we shall introduce 2
extra symmetries with nilpotent generator of the complete action
(\ref{completeaction}). We first determine the BRST charge in
functional form. We shall see that it remains unchanged compared to
the Yang-Mills case (\ref{br9}). The Noether current corresponding
to the BRST transformation (\ref{ourbrst}) and action (\ref{last1})
is given by
\begin{eqnarray}\label{z17}
\mathcal{J}_0^\mu &=&F^{\mu\nu,a}\p_\nu c^a-b^a\p^\mu
c^a-\frac{im}{2}\left(B-\overline{B}\right)^{\mu\nu,a}\p_\nu c^a\,,
\end{eqnarray}
which leads to the BRST charge
\begin{eqnarray}\label{z18}
    \mathcal Q_0&=&\int d^3x\left(F^{0i}\p_i c^a-b^a\p^0c^a-\frac{im}{2}\left(B-\overline{B}\right)^{0i,a}\p_i c^a\right)\nonumber\\
&=&\int d^3x\left(c^a\p^0 b^a -b^a\p^0c^a\right)\,,
\end{eqnarray}
where we invoked the equation of motion
\begin{equation}\label{z19}
    \p_\mu F^{\mu\nu}=\p^\nu
b^a+\frac{im}{2}\p_\mu\left(B-\overline{B}\right)^{\mu\nu,a}\,.
\end{equation}
For what concerns the Faddeev-Popov ghosts $c$ and $\oc$, it is
immediately seen from the action (\ref{last1}) that their
quantization remains unchanged compared to the Yang-Mills case,
given in (\ref{br1}), (\ref{br5}) and (\ref{br6}). Therefore, since
(\ref{z18}) must be time independent as a conserved charge, we
already infer that $\mathcal{Q}_0$ will only act nontrivially  on
massless excitations. This shall be confirmed later once we have
found the excitations belonging to the $b^a$-field (see section 6).

\sect{A further reduction of the physical subspace} In the following
sections, we shall make use of a cohomological result
\cite{Piguet:1995er}, summarized here.

\noindent \textbf{Doublet theorem} Consider a transformation
$\delta$ with the property that
\begin{eqnarray}
\begin{array}{lll}
  \delta u_i = v_i \,,&\quad&  \delta u_i^\prime = v_i^\prime\,,\\
  \delta v_i = 0\,,&\quad & \delta v_i^\prime = 0\,,
\end{array}
\end{eqnarray}
with $u_i,v_i^\prime$ commuting and $u_i^\prime,v_i$ anticommuting
quantities. We call $(u_i,v_i)$ and $(u_i^\prime,v_i^\prime)$
$\delta$-doublets.

\noindent Then it is a trivial exercise to show that $\delta$ is a
nilpotent transformation. Moreover, $(u_i,v_i)$ and
$(u_i^\prime,v_i^\prime)$ appear trivially in the
$\delta$-cohomology. This can be proven \cite{Piguet:1995er} by
introducing the ``counting'' operator
\begin{equation}\label{com1}
    \mathcal{P}=\int d^4x \left(u_i\frac{\delta}{\delta u_i}+v_i\frac{\delta}{\delta
v_i}+u_i^\prime\frac{\delta}{\delta
u_i^\prime}+v_i^\prime\frac{\delta}{\delta v_i^\prime}\right)\,,
\end{equation}
and the operator
\begin{equation}\label{com2}
    \mathcal{A}=\int d^4x \left(u_i\frac{\delta}{\delta v_i}+u_i^\prime\frac{\delta}{\delta
v_i^\prime}\right)\,,
\end{equation}
such that
\begin{equation}\label{com3}
    \mathcal{P}=\left\{\delta,\mathcal{A}\right\}\qquad\qquad
\left[\mathcal{P},\delta\right]=0\,.
\end{equation}
Assuming that $\delta X=0$, we can expand $X$ in an mutual
eigenbasis of the commuting Hermitian operators $\mathcal{P}$ and
$\delta$. It is then quite easy to show that
\begin{equation}\label{com4}
    X=X_0+ \delta Y\,,
\end{equation}
whereby $\mathcal{P}X_0=0$, i.e. the cohomology of $\delta$ does not
depend on $(u_i,v_i)$ and $(u_i^\prime,v_i^\prime)$.

\noindent In the paper \cite{Capri:2006ne}, we already noted the
equivalence between the action (\ref{completeaction}) for $m\equiv0$
and conventional Yang-Mills theories, when quantized using the same
gauge fixing. The physical content of both theories should be the
same, in casu the unitarity should be satisfied. This can be shown
in the following way. The action (\ref{completeaction}) enjoys a
supersymmetry when $m\equiv0$, generated by the nilpotent
transformation \cite{Capri:2006ne}
\begin{eqnarray}\label{supersym}
\begin{array}{lll}
  \delta_s B_{\mu\nu}^a = G_{\mu\nu}^a\,, & \quad & \delta_s \overline{G}_{\mu\nu}^a = \overline{B}_{\mu\nu}^a\,,\\
  \delta_s G_{\mu\nu}^a = 0\,, & \quad & \delta_s
\oB_{\mu\nu}^a=0\,.
\end{array}
\end{eqnarray}
So, the \textbf{Doublet theorem} applies, and we conclude that the
excitations belonging to the extra fields will not belong to the
physical subspace.

\noindent In the case that $m\neq0$, the supersymmetry
(\ref{supersym}) is broken due to the terms $\sim(B-\oB)$. In a
matter of speaking, the symmetry is only broken by terms
$\sim(B-\oB)$, and not by terms $\sim(B+\oB)$. Therefore, we might
expect that some trace of the supersymmetry $\delta_s$ might survive
after all. We shall first explore this possibility. We decompose the
fields $G_{\mu\nu}^a$ and $\oG_{\mu\nu}^a$ in their ``electric'' and
``magnetic'' part
\begin{eqnarray}\label{decom2}
\begin{array}{lll}
  \al ^{i,a}=\frac{1}{2}\varepsilon^{ijk}G_{jk}^a\,, & \quad & \overline{\al}^{i,a}=\frac{1}{2}\varepsilon^{ijk}\overline{G}_{jk}^a\,,\\
  \be^{i,a}=G^{oi,a}\,, & \quad &
\overline{\be}^{i,a}=\overline{G}^{oi,a}\,.
\end{array}
\end{eqnarray}
Consequently, one finds
\begin{equation}\label{use1}
    -\frac{1}{4}\overline{G}_{\mu\nu}^a\p^2
    G^{\mu\nu,a}=\frac{1}{2}\overline{\be}^{i,a}\p^2 \be^{i,a}-\frac{1}{2}\overline{\al}^{i,a}\p^2
    \al^{i,a}\,.
\end{equation}
Since $G$ and $\oG$ always appear in the ``product'' combination
$\oG_{\mu\nu}^a \mathcal{O}^{ab} G^{\mu\nu,b}$, only the structures
$\overline{\be}\mathcal{O}\be$ and $\overline{\al}\mathcal{O}\al$
will appear. Having a look at the complete action
(\ref{completeaction}), it is clear that also in the interacting
theory, these are the only possibly appearing structures.

\noindent It is convenient to also decompose the $B$ and
$\oB$-fields in their electric and magnetic counterparts
\begin{eqnarray}\label{last2}
\begin{array}{lll}
  \rho ^{i,a}=\frac{1}{2}\varepsilon^{ijk}B_{jk}^a\,, & \quad &
\overline{\rho}^{i,a}=\frac{1}{2}\varepsilon^{ijk}\overline{B}_{jk}^a\,,\\
  \sigma^{i,a}=B^{oi,a}\,, & \quad &
\overline{\sigma}^{i,a}=\overline{B}^{oi,a}\;.
\end{array}
\end{eqnarray}
in which case the (quadratic) action becomes
\begin{eqnarray}\label{last3}
  S_0&=&\int d^4x\left[-\frac{1}{4}\left(\p_\mu A_\nu^a-\p_\nu A_\mu^a\right)^2-\frac{im}{2}(\sigma-\overline{\sigma})^{i,a}\left(\p^0 A^{i a}-\p^i A^{0 a}\right)
  \right.\nonumber\\&+&\left.\frac{im}{2}\varepsilon^{ijk}(\rho-\overline{\rho})^{k,a}\left(\p^i
A^{j a}\right)-\frac{1}{2} \overline{\sigma}^{i ,a}\p^2 \sigma^{i,
a}+\frac{1}{2} \overline{\rho}^{i ,a}\p^2
\rho^{i, a}+\frac{1}{2}\overline{\beta}^{i,a}\p^2 \beta^{i,a}-\frac{1}{2}\overline{\al}^{i,a}\p^2 \al^{i,a}\right.\nonumber\\&-&\left.\frac{3}{4}%
m^{2}\lambda _{1}\left(
-\overline{\sigma}^{i,a}\sigma^{i,a}+\overline{\rho}^{i,a}\rho^{i,a}+\overline{\beta}^{i,a}
\beta^{i,a}-\overline{\al}^{i,a} \al^{i,a}\right)
\right.\nonumber\\&+&\left.m^{2}\frac{\lambda
_{3}}{8}\left(\overline{\sigma}^{i,a}\sigma^{i,a}-\overline{\rho}^{i,a}\rho^{i,a}\right)
+m^{2}\frac{\lambda _{3}}{16}\left(
-\overline{\sigma}^{i,a}\overline{\sigma}^{i,a}-\sigma^{i,a}\sigma^{i,a}+\overline{\rho}^{i,a}\overline{\rho}^{i,a}
+\rho^{i,a}\rho^{i,a}\right) \right.\nonumber\\&+&\left.b^{a}%
\partial _{\mu }A^{\mu a}+\overline{c}^{a}\p^2
c^{a}+\frac{\xi }{2}b^{a}b^{a}\right]\,.
\end{eqnarray}
For further convenience, we shall exchange $(\sigma,
\overline{\sigma})$ and $(\rho,\overline{\rho})$ for their real and
imaginary parts via
\begin{eqnarray}\label{last4}
\begin{array}{lll}
  K^{i,a}=\frac{1}{2}(\rho^{i,a}+\overline{\rho}^{i,a})\,, & \quad & M^{i,a}=\frac{1}{2}(\sigma^{i,a}+\overline{\sigma}^{i,a})\,,\\
  L^{i,a}=\frac{1}{2i}(\rho^{i,a}-\overline{\rho}^{i,a})\,, & \quad
&N^{i,a}=\frac{1}{2i}(\sigma^{i,a}-\overline{\sigma}^{i,a})\,,
\end{array}
\end{eqnarray}
yielding
\begin{eqnarray}\label{last5}
  S_0&=&\int d^4x\left[-\frac{1}{4}\left(\p_\mu A_\nu^a-\p_\nu A_\mu^a\right)^2+mN^{i,a}\left(\p^0 A^{i a}-\p^i A^{0 a}\right)
-m\varepsilon^{ijk}L^{k,a}\left(\p^i A^{j a}\right)
  \right.\nonumber\\&-&\left.\frac{1}{2} M^{i
,a}\p^2 M^{i, a}-\frac{1}{2} N^{i ,a}\p^2 N^{i, a}+\frac{1}{2} K^{i
,a}\p^2 K^{i, a}\right.\nonumber\\&+&\left.\frac{1}{2} L^{i ,a}\p^2
L^{i,
a}+\frac{1}{2}\overline{\beta}^{i,a}\p^2 \beta^{i,a}-\frac{1}{2}\overline{\al}^{i,a}\p^2 \al^{i,a}\right.\nonumber\\&-&\left.\frac{3}{4}%
m^{2}\lambda _{1}\left(
-M^{i,a}M^{i,a}-N^{i,a}N^{i,a}+K^{i,a}K^{i,a}+L^{i,a}L^{i,a}+\overline{\beta}^{i,a}
\beta^{i,a}-\overline{\al}^{i,a} \al^{i,a}\right)
\right.\nonumber\\&+&\left. m^{2}\frac{\lambda
_{3}}{4}\left(N^{i,a}N^{i,a}-L^{i,a}L^{i,a}\right)+b^{a}%
\partial _{\mu }A^{\mu a}+\overline{c}^{a}\p^2
c^{a}+\frac{\xi }{2}b^{a}b^{a}\right]\,.
\end{eqnarray}
As a final step, we introduce the fields
\begin{equation}\label{last5a}
\chi^{\pm,i,a}=K^{i,a}\pm M^{i,a}\,,
\end{equation}
to write
\begin{eqnarray}\label{last5b}
  S_0&=&\int d^4x\left[-\frac{1}{4}\left(\p_\mu A_\nu^a-\p_\nu A_\mu^a\right)^2+mN^{i,a}\left(\p^0 A^{i a}-\p^i A^{0 a}\right)
-m\varepsilon^{ijk}L^{k,a}\left(\p^i A^{j a}\right)
  \right.\nonumber\\&+&\left.\frac{1}{2} \chi^{+,i
,a}\p^2 \chi^{-, i, a}-\frac{1}{2} N^{i ,a}\p^2 N^{i, a}+\frac{1}{2}
L^{i ,a}\p^2 L^{i,
a}+\frac{1}{2}\overline{\beta}^{i,a}\p^2 \beta^{i,a}-\frac{1}{2}\overline{\al}^{i,a}\p^2 \al^{i,a}\right.\nonumber\\&-&\left.\frac{3}{4}%
m^{2}\lambda _{1}\left(
\chi^{+,i,a}\chi^{-,i,a}-N^{i,a}N^{i,a}+L^{i,a}L^{i,a}+\overline{\beta}^{i,a}
\beta^{i,a}-\overline{\al}^{i,a} \al^{i,a}\right)
\right.\nonumber\\&+&\left. m^{2}\frac{\lambda
_{3}}{4}\left(N^{i,a}N^{i,a}-L^{i,a}L^{i,a}\right)+b^{a}%
\partial _{\mu }A^{\mu a}+\overline{c}^{a}\p^2
c^{a}+\frac{\xi }{2}b^{a}b^{a}\right]\,.
\end{eqnarray}
We introduce the following transformations
\begin{eqnarray}\label{last5c}
\begin{array}{lll}
  \Delta_\al \overline{\al}^{i,a} = \chi^{+,i,a}\,, & \quad & \Delta_\al \chi^{-,i,a}=\al^{i,a}\,, \\
 \Delta_\al\chi^{+,i,a}=0 \,, & \quad & \Delta_\al\al^{i,a}=0\,,
\end{array}
\end{eqnarray}
and
\begin{eqnarray}\label{last5d}
\begin{array}{lll}
\Delta_\beta \overline{\beta}^{i,a} = \chi^{+,i,a}\,,&\quad&\Delta_\beta \chi^{-,i,a}=-\beta^{i,a}\,, \\
\Delta_\beta\chi^{+,i,a}=0\,,&\quad&\Delta_\beta\beta^{i,a}=0\,.
\end{array}
\end{eqnarray}
Clearly, these transformations define a symmetry of the free action
$S_0$ (\ref{last5b}). However, these will also generate a symmetry
of the full action (\ref{completeaction}). One notices that only the
3rd, 4th and 6th term of (\ref{completeactionb}) will give rise to
contributions in $\chi^{\pm}$. Taking a closer look at these terms,
it is quite easily seen that these contributions will always be of
the type
\begin{equation}\label{last5e}
    \chi^{+,i,a}\mathcal{O}^{ab}\chi^{-,i,b}+\overline{\beta}^{i,a}\mathcal{O}^{ab}\beta^{i,b}-\overline{\al}^{i,a}\mathcal{O}^{ab}\al^{i,b}\,,
\end{equation}
with
\begin{equation}\label{last5ebis}
    \mathcal{O}^{ab}=\left\{\begin{array}{l}
                       \delta^{ab} \\
D_{\mu}^{ac}D^{\mu,cb}=\delta^{ab}\p^2-gf^{abd}\left(\p^\mu
A_{\mu}^d+2A_\mu^d\p^\mu\right)+g^2f^{acd}f^{cbe}A_\mu^dA^{\mu,e}
                     \end{array}\right.
\end{equation}
and one can check that (\ref{last5e}) vanishes when
$\Delta_{\al,\beta}$ is applied to it.

\noindent It is also readily derived that
\begin{equation}\label{last5f}
    \left\{\Delta_\al,\Delta_\beta\right\}=0\,,
\end{equation}
while it also holds that
\begin{equation}\label{last5g}
    \left\{\Delta_\al,s\right\}=0=\left\{\Delta_\beta,s\right\}\,,
\end{equation}
since the BRST operator $s$ acting on the new fields can be read off
from (\ref{brst3}), (\ref{decom2}), (\ref{last4}) and (\ref{last5a})
to be
\begin{eqnarray}\label{last5g}
    s\Omega^{a,i}&=&gf^{abc}c^b\Omega^{c,i}\,,\nonumber\\
\Omega^{a,i}&\in&\left\{\alpha^{i,a},\overline{\al}^{i,a},\beta^{i,a},\overline{\beta}^{i,a},\chi^{+,i,a},\chi^{-,i,a},L^{i,a},N^{i,a}\right\}\,.
\end{eqnarray}
Recapitulating, we have found 2 symmetries of the action
(\ref{completeaction}) which are generated by the nilpotent
generators (\ref{last5c}) and (\ref{last5d}). Moreover, the fields
$(\overline{\al}^{i,a},\chi^{+,i,a})$ and $(\chi^{-,i,a},\al^{i,a})$
form $\Delta_\al$ doublets, so we can be assured that these fields
decouple from the physical sector by applying the \textbf{Doublet
theorem}. Moreover, we can equally well deploy the \textbf{Doublet
theorem} on $\Delta_\beta$ to also remove $\beta$ and
$\overline{\beta}$ from the physical subspace. The intersection
$\mbox{cohom}(\Delta_{\al})\cap\mbox{cohom}(\Delta_{\beta})$ is thus
independent of $\alpha^{i,a}$, $\overline{\al}^{i,a}$, $
\beta^{i,a}$, $ \overline{\beta}^{i,a}$, $\chi^{+,i,a}$ and
$\chi^{-,i,a}$.

\noindent Consequently, we can conclude that all degrees of freedom
corresponding to the extra ghost fields $G_{\mu\nu}^a$ and
$\overline{G}_{\mu\nu}^a$ as well as the extra bosonic degrees of
freedom corresponding to $B_{\mu\nu}^a+\overline{B}_{\mu\nu}^a$ are
decoupled from the physical sector.

\noindent For the fields $N^{i,a}$ and $L^{i,a}$, corresponding to
the degrees of freedom in $B_{\mu\nu}^a-\overline{B}_{\mu\nu}^a$,
the analysis is less quickly performed, as these are coupled to the
original gluon field $A_\mu^a$ in a nontrivial way. In the next
sections, we will have a look at this problem. As we did not analyze
yet the space annihilated by the free BRST charge, we may expect
that certain degrees of freedom will also be killed when this
subspace is considered.

\sect{The classical equations of motion and the Fourier
decomposition of the fields} As a next step, we must quantize our
model. Before turning to the conjugate momenta and quantization
rules, it is advisable to have a look at the classical equations of
motion in the new variables and the corresponding Fourier
decompositions of their solutions. Due to the mixing between the
fields, their Fourier coefficients will not all be independent.

\subsection{Classical equations of motion}
 The free classical equation of motions are\footnote{We shall skip again the global color index.}
\begin{eqnarray}
\label{z1a}b+\p^\mu A_\mu&=&0\,,\\
     \label{z1b}\p^2A_0-m\p_i N^{i}&=&0\,,\\
\label{z1c}(\p^2+\widetilde{m}^2)N^{i}-m(\p^0
A^{i}-\p^iA^{0})&=&0 \,,\\
\label{z1e}-(\p^2+\widetilde{m}^2)L^{i}-m\varepsilon^{ijk}\p_j
A^k&=&0\,,\\
\label{z1d}\p^2 A^{i}+m\p_0 N^{i}-m\varepsilon^{ijk}\p_j
L^{k}&=&0\,,
\end{eqnarray}
where we set
\begin{equation}
\widetilde{m}^2=-\frac{3}{2}m^2\lambda_1-\frac{1}{2}m^2\lambda_3\,.
\end{equation}
Since these equations are coupled, the quantization of the fields is
not straightforward. \subsubsection{Intermezzo: multipole
fields}After some manipulation with the equations
(\ref{z1a})-(\ref{z1d}), we derive that
\begin{eqnarray}\label{z14}
    \p^2\left(\p^2+m^2+\widetilde{m}^2\right)(\p^2A_0)&=&0\,,\nonumber\\
\p^2\left(\p^2+m^2+\widetilde{m}^2\right)(\p_iN^i)&=&0\,,\nonumber\\
\left(\p^2+\widetilde{m}^2\right)(\p_iL^i)&=&0\,,\nonumber\\
    \p^2\left(\p^2+m^2+\widetilde{m}^2\right)\p^2(\p_i A^i)&=&0\,.
\end{eqnarray}
These (induced) equations of motion are of higher order in the
derivatives. In the literature, such fields are known as
``multipole'' fields, and their quantization is indeed more involved
than in the well known Klein-Gordon case. The problem of dipole
fields also appears in the supergravity context
\cite{Ferrara:1977mv}.

\noindent Consider e.g. the following toy model of a higher
derivative action \cite{Ferrara:1977mv}
\begin{equation}\label{mu2}
    S=-\frac{1}{2}\phi\frac{\p^2}{m^2} \p^2\phi\,,
\end{equation}
with classical equation of motion
\begin{equation}\label{mu3}
    \p^2 \p^2\phi=0\,.
\end{equation}
It is not clear how to quantize the model (\ref{mu2}) since higher
order derivatives occur. We can rewrite (\ref{mu2}) as
\begin{equation}\label{mu4}
    S=-\psi \p^2\phi+\frac{1}{2}m^2\psi^2\,,
\end{equation}
by introducing an auxiliary field $\psi$. It is important to notice
that the derivates now occur at most quadratically, making the
action (\ref{mu4}) suitable for canonical quantization. In fact,
this is the case we are investigating. Our action (\ref{last5}) is
indeed already at most quadratical in the derivatives.

\noindent The equations of motion associated to (\ref{mu4}) are
\begin{eqnarray}\label{mu5}
  \p^2 \phi &=& m^2\psi\,, \nonumber\\
   \p^2 \psi &=&0\,.
\end{eqnarray}
Let us turn to Fourier space. We propose the solution
\begin{eqnarray}\label{mu6}
  \psi(x) &=& \int\frac{d^3q}{(2\pi)^3}\widehat{\psi}(\vq,t)e^{i\vec{q}\vec{x}}+\mbox{h.c.}\,,  \\
  \phi(x) &=&
\int\frac{d^3q}{(2\pi)^3}\widehat{\phi}(\vq,t)e^{i\vec{q}\vec{x}}+\mbox{h.c.}\,.
\end{eqnarray}
Plugging (\ref{mu6}) in (\ref{mu5}), the following differential
equations in the time $t\equiv x_0$ arise
\begin{eqnarray}
  \label{mu7a}(\p_0^2+\vec{q}^2)\widehat{\phi}(\vec{q},t)&=&m^2\widehat{\psi}(\vq,t)\,,\nonumber\\
\label{mu7b}(\p_0^2+\vec{q}^2)\widehat{\psi}(\vec{q},t)&=&0\,.
\end{eqnarray}
We can set
\begin{equation}
\widehat{\psi}(\vec{q},t)=a(\vq)e^{-iq_0t}\,,\qquad q_0=|\vec{q}|\,,
\end{equation}
hence the equation (\ref{mu7a}) is satisfied when
\begin{equation}
\widehat{\phi}(\vec{q},t)=\underbrace{b(\vq)e^{-iq_0t}}_{\mbox{{\tiny
solution
hom.eq.}}}+\frac{m^2}{4\vec{q}^2}(1+2i|\vq|t)a(\vq)e^{-iq_0t}\,,\qquad
q_0=|\vec{q}|\,.
\end{equation}
According to \cite{Ferrara:1977mv}, terms linear in time $t$ are
typical for dipoles. As we shall soon see, terms $\propto t$ shall
also appear in our massive gauge model. One can then continue by
quantizing the model. The necessary commutation relations are of the
type
\begin{equation}\label{mu10}
    \left[a(q),b^\dagger(k)\right]=\left[b(q),a^\dagger(k)\right]=\frac{1}{2k_0}
\delta^{(3)}(\vec{k}-\vec{q})\,.
\end{equation}
Similar, although a little more complicated techniques and
relations, will occur when we try to solve the classical equations
of motions of our model in Fourier space.

\subsubsection{Solving the equations of motion} Let us now try to solve the equations
(\ref{z1a})-(\ref{z1d}) in Fourier space. We propose
\begin{eqnarray}\label{mu11}
    A_0(x)&=&\int\frac{d^3q}{(2\pi)^3}\widehat{A}_0(\vq,t)e^{i\vq\vec{x}}+\mbox{h.c.}\,,\nonumber\\
A^i(x)&=&\int\frac{d^3q}{(2\pi)^3}\sum_{m=1}^3\left(\widehat{A}_m(\vq,t)\varepsilon^{(m),i}(\vec{q})e^{i\vec{q}\vec{x}}\right)+\mbox{h.c.}\,,\nonumber\\
    N^i(x)&=&\int\frac{d^3q}{(2\pi)^3}\sum_{m=1}^3\left(\widehat{N}_m(\vq,t)\varepsilon^{(m),i}(\vec{q})e^{i\vq\vec{x}}\right)+\mbox{h.c.}\,,\nonumber\\
   L^i(x)&=&\int\frac{d^3q}{(2\pi)^3}\sum_{m=1}^3\left(\widehat{L}_m(\vq,t)\varepsilon^{(m),i}(\vec{q})e^{i\vq\vec{x}}\right)+\mbox{h.c.}\,,\nonumber\\
b(x)&=&\int\frac{d^3q}{(2\pi)^3} \left(-\p_0
\widehat{A}_0(\vq,t)-i|\vec{q}|\widehat{A}_3(\vq,t)\right)e^{i\vq\vec{x}}+\mbox{h.c.}\,.
\end{eqnarray}
The next step is to derive the equations for the Fourier
coefficients. Plugging (\ref{mu11}) into (\ref{z1b}), we derive
\begin{equation}\label{mu12}
(\p_0^2+\vq^2)\widehat{A}_0=mi|\vec{q}|\widehat{N}_3\,,
\end{equation}
while (\ref{z1c}) yields
\begin{eqnarray}
\label{mu13a}(\p_0^2+\vq^2+\widetilde{m}^2)\widehat{N}_3&=&m\left(\p_0 \widehat{A}_3+i|\vec{q}|\widehat{A}_0\right)\,,\\
\label{mu13b}(\p_0^2+\vq^2+\widetilde{m}^2)\widehat{N}_{1,2}&=&m\p_0
\widehat{A}_{1,2}\,.
\end{eqnarray}
(\ref{z1e}) results in
\begin{eqnarray}
\label{mu14}-(\p_0^2+\vq^2+\widetilde{m}^2)\sum_{m=1}^{3}\widehat{L}_m\varepsilon^{(m),i}&=&m\varepsilon^{ijk}iq^j\sum_{m=1}^3\varepsilon^{(m),k}\widehat{A}_m\,.
\end{eqnarray}
Multiplying with $\varepsilon^{(n),i}$ , summing over $i$ and using
the orthonormality, we find
\begin{eqnarray}
\label{mu15}-(\p_0^2+\vq^2+\widetilde{m}^2)\widehat{L}_n&=&-im\sum_{m=1}^3
\widehat{A}_m
q^j\left(\varepsilon^{(n)}\times\varepsilon^{(m)}\right)^j\,.
\end{eqnarray}
Taking $n=3$, we know that
$\varepsilon^{(m)}\times\varepsilon^{(n)}$ will only survive for
$m=1,2$, however then the cross product is
$\propto\varepsilon^{(2),(1)}$ and thus orthogonal to $\vec{q}$,
meaning that
\begin{eqnarray}
\label{mu16} (\p_0^2+q^2+\widetilde{m}^2)\widehat{L}_3&=&0\,.
\end{eqnarray}
With $n=1,2$, (\ref{mu15}) is valid if
\begin{eqnarray}
\label{mu17a} -(\p_0^2+q^2+\widetilde{m}^2)\widehat{L}_1&=&-im|\vec{q}|\widehat{A}_2\,,\\
\label{mu17b}-(\p_0^2+q^2+\widetilde{m}^2)\widehat{L}_2&=&im|\vec{q}|\widehat{A}_1\,.
\end{eqnarray}
Analogous manipulations on (\ref{z1d}) give
\begin{eqnarray}
\label{mu18a} (\p_0^2+q^2)\widehat{A}_3+m\p_0\widehat{N}_3&=&0\,,\\
\label{mu18b}(\p_0^2+q^2)\widehat{A}_1+m\p_0 \widehat{N}_1&=&-im|\vec{q}|\widehat{L}_2\,,\\
\label{mu18c}(\p_0^2+q^2)\widehat{A}_2+m\p_0
\widehat{N}_2&=&im|\vec{q}|\widehat{L}_1\,.
\end{eqnarray}
Let us now try to find a sensible solution to the previous
differential equations in $t$.

\noindent We first concentrate on the independent subset of
equations (\ref{mu12}), (\ref{mu13a}) and (\ref{mu18a}). We can
decouple these differential equations by passing to higher order
differential equations. These induced equations read
\begin{eqnarray}\label{mu19}
    (\p_0^2+\vq^2)^2(\p_0^2+\vec{q}^2+M^2)\widehat{A}_{0,3}&=&0\,,\nonumber\\
(\p_0^2+\vq^2)(\p_0^2+\vec{q}^2+M^2)\widehat{N}_{3}&=&0\,,
\end{eqnarray}
where we defined
\begin{equation}
M^2=m^2+\wm^2\,.
\end{equation}
We solve these equations by\footnote{We did not write explicitly the
$\mbox{h.c.}$ part.}
\begin{eqnarray}\label{mu20}
\widehat{A}_0(\vq,t)&=&a_0(\vec{q})e^{-i|\vec{q}|t}+t\widetilde{a}_0(\vec{q})e^{-i|\vec{q}|t}+\widehat{a}_0(\vec{q})e^{-i\sqrt{\vq^2+M^2}t}\,,\nonumber\\
\widehat{A}_3(\vq,t)&=&a_3(\vec{q})e^{-i|\vec{q}|t}+t\widetilde{a}_3(\vec{q})e^{-i|\vec{q}|t}+\widehat{a}_3(\vec{q})e^{-i\sqrt{\vq^2+M^2}t}\,,\nonumber\\
\widehat{N}_3(\vq,t)&=&n_3(\vec{q})e^{-i|\vec{q}|t}+\widehat{n}_3(\vec{q})e^{-i\sqrt{\vq^2+M^2}t}\,.
\end{eqnarray}
For consistency, we must impose the original set (\ref{mu12}),
(\ref{mu13a}) and (\ref{mu18a}) again, and identify the coefficients
of the various $t$-dependent functions ($e^{-i|\vec{q}|t}$,
$te^{-i|\vec{q}|t}$ and $e^{-i\sqrt{\vq^2+M^2}t}$). The following
equations come out as independent ones
\begin{eqnarray}\label{mu21}
  -M^2\widehat{a}_0&=& mi|\vec{q}|\widehat{n}_3 \,,\nonumber\\
   -2\widetilde{a}_0&=&mn_3\,,\nonumber\\
    \widetilde{a}_3 &=& \widetilde{a}_0\,,\nonumber\\
-mi|\vq|a_3+m\widetilde{a}_3+mi|\vq|a_0&=&\wm^2n_3\,, \nonumber\\
-M^2\widehat{a}_3&=&mi\sqrt{M^2+\vq^2}\widehat{n}_3\,,
\end{eqnarray}
implying that there are only 3 independent coefficients left of the
original 8 in (\ref{mu20}).

\noindent The next step is to analyze the remaining equations.
(\ref{mu16}) immediately gives
\begin{equation}\label{mu22}
    \widehat{L}_3(\vq,t)=\widetilde{\ell}_3(\vec{q})e^{-i\sqrt{\vq^2+\wm^2}t}\,.
\end{equation}
The last independent set consists of (\ref{mu13b}),(\ref{mu17a}),
(\ref{mu17b}), (\ref{mu18b}) \& (\ref{mu18c}). Applying the same
trick as before, we deduce
\begin{eqnarray}\label{mu23}
    (\p_0^2+\vq^2)(\p_0^2+\vec{q}^2+M^2)\widehat{A}_{1,2}&=&0\,,\nonumber\\
(\p_0^2+\vq^2)(\p_0^2+\vec{q}^2+\wm^2)(\p_0^2+\vec{q}^2+M^2)\widehat{N}_{1,2}&=&0\,,\nonumber\\
(\p_0^2+\vq^2)(\p_0^2+\vec{q}^2+\wm^2)(\p_0^2+\vec{q}^2+M^2)\widehat{L}_{1,2}&=&0\,,
\end{eqnarray}
with corresponding solutions\footnote{We assume that $\wm^2\neq0$,
otherwise we should adapt the analysis.}
\begin{eqnarray}\label{mu24}
\widehat{A}_1(\vq,t)&=&a_1(\vec{q})e^{-i|\vec{q}|t}+\widehat{a}_1(\vec{q})e^{-i\sqrt{\vq^2+M^2}t}\,,\nonumber\\
\widehat{A}_2(\vq,t)&=&a_2(\vec{q})e^{-i|\vec{q}|t}+\widehat{a}_2(\vec{q})e^{-i\sqrt{\vq^2+M^2}t}\,,\nonumber\\
\widehat{L}_1(\vq,t)&=&\ell_1(\vec{q})e^{-i|\vec{q}|t}+\widetilde{\ell}_1(\vec{q})e^{-i\sqrt{\vq^2+\wm^2}t}+\widehat{\ell}_1(\vec{q})e^{-i\sqrt{\vq^2+M^2}t}\,,\nonumber\\
\widehat{L}_2(\vq,t)&=&\ell_2(\vec{q})e^{-i|\vec{q}|t}+\widetilde{\ell}_2(\vec{q})e^{-i\sqrt{\vq^2+\wm^2}t}+\widehat{\ell}_2(\vec{q})e^{-i\sqrt{\vq^2+M^2}t}\,,\nonumber\\
\widehat{N}_1(\vq,t)&=&n_1(\vec{q})e^{-i|\vec{q}|t}+\widetilde{n}_1(\vec{q})e^{-i\sqrt{\vq^2+\wm^2}t}+\widehat{n}_1(\vec{q})e^{-i\sqrt{\vq^2+M^2}t}\,,\nonumber\\
\widehat{N}_2(\vq,t)&=&n_2(\vec{q})e^{-i|\vec{q}|t}+\widetilde{n}_2(\vec{q})e^{-i\sqrt{\vq^2+\wm^2}t}+\widehat{n}_2(\vec{q})e^{-i\sqrt{\vq^2+M^2}t}\,.
\end{eqnarray}
Consistency requires
\begin{eqnarray}\label{mu1000}
 \begin{array}{lll}
   \wm^2 n_{1} = -im|\vq| a_1\;, & \quad&\wm^2 n_{2} = -im|\vq| a_2\,,\\
m \widehat{n}_{1} = i\sqrt{\vq^2+M^2} \widehat{a}_1\;,  &\quad&   m \widehat{n}_{2} = i\sqrt{\vq^2+M^2} \widehat{a}_2\,,  \\
   m \widehat{\ell}_1= -i|\vq|\widehat{a}_2\;,  &\quad&  m \widehat{\ell}_2= i|\vq|\widehat{a}_1\,,\\
   n_1=\ell_2\;, &\quad& n_2=-\ell_1\,,\\
  \sqrt{\vq^2+\wm^2}\widetilde{n}_1=|\vq|\widetilde{\ell}_2\;,
&\quad& \sqrt{q^2+\wm^2}\widetilde{n}_2=-|\vq|\widetilde{\ell}_1\,.
 \end{array}
\end{eqnarray}
This reduces the number of independent Fourier coefficients in
(\ref{mu24}) from 16 to 6.

\noindent Summarizing, we have 3+1+6=10 independent Fourier
coefficients left (plus their hermitian conjugates). Without loss of
generality, we choose to work with $n_3(\vq)$, $a_1(\vq)$,
$a_2(\vq)$, $\widehat{a}_1(\vq)$, $\widehat{a}_2(\vq)$,
$\widehat{n}_3(\vq)$, $\widehat{\ell}_3(\vq)$,
$\widetilde{\ell}_1(\vq)$, $\widetilde{\ell}_2(\vq)$ and $a_0(\vq)$.

\subsection{The conjugate momenta}
In order to quantize the theory, we need the conjugate momenta. We
shall only be concerned about the degrees of freedom hidden in the
bosonic fields. We do not care about the ghosts for the moment.

\noindent Making use of the action (\ref{last5}), we derive the
desired conjugate momenta,
\begin{eqnarray}\label{z14}
  \pi_{i}^a &=& \frac{\delta\cal L}{\delta (\p_0 A^{i,a})}=-F_{0i}^a-mN_i^a\,,\nonumber \\
\pi_{N,i}^a&=&\frac{\delta\cal L}{\delta (\p_0 N^{i,a})}=-\p_{0}N_{i}^a+mA_i^a\,,\nonumber \\
\pi_{L,i}^a&=&\frac{\delta\cal L}{\delta (\p_0 L^{i,a})}=\p_0 L_i^a\,,\nonumber\\
  \pi^{0,a} &=& b^a\equiv -\p^{\mu} A_{\mu}^a\,.
\end{eqnarray}
For later use, let us compute the the multiplier $b$ of
(\ref{mu11}), which is actually given by
\begin{eqnarray}\label{mu30}
    b(x)&=&\frac{M^2}{m}\int\frac{d^3q}{(2\pi)^3}
n_3(\vq)e^{-i|\vq|t}e^{i\vq\vec{x}}+\mbox{h.c.}\,.
\end{eqnarray}
Let us now quantize the model. Naively, the brackets we would impose
are\footnote{See also the Appendix.}
\begin{eqnarray}\label{q1}
    \left[A_i(\vec{x},t),\pi_{j}(\vec{y},t)\right]&=&-i\delta_{ij}\delta^{(3)}(\vec{x}-\vec{y})\,,\nonumber\\
    \left[A_0(\vec{x},t),\pi^{0}(\vec{y},t)\right]&=&i\delta^{(3)}(\vec{x}-\vec{y})\,,\nonumber\\
    \Bigl\lbrace c(\vec{x},t),\pi_c(\vec{y},t)\Bigr\rbrace&=&i\delta^{(3)}(\vec{x}-\vec{y})\,,\nonumber\\
\Bigl\lbrace\occ(\vec{x},t),\pi_{\occ}(\vec{y},t)\Bigr\rbrace&=&i\delta^{(3)}(\vec{x}-\vec{y})\,,\nonumber\\
\left[N_i(\vec{x},t),
\pi^N_j(\vec{y},t)\right]&=&-i\delta_{ij}\delta^{(3)}(\vec{x}-\vec{y})\,,\nonumber\\
\left[L_i(\vec{x},t),
\pi^L_j(\vec{y},t)\right]&=&i\delta_{ij}\delta^{(3)}(\vec{x}-\vec{y})\,.
\end{eqnarray}
However, these brackets will be not necessarily correct, due to the
fact that multiple relations exist between the different field and
conjugate momenta configurations, something which is clearly visible
from the relations between the Fourier coefficients (\ref{mu21}) and
(\ref{mu1000}). We postpone the actual discussion of the
quantization to section 7, where we shall explain an alternative way
to fix the commutation relations in an appropriate fashion.

\noindent An additional complication that we should have a look at
is the precise form of the free Hamiltonian $\mathcal{H}_0$. In
order to have a well defined Fock state space describing physical
particles, the states we are considering, which are defined by
acting with the creation operators on the vacuum, should be energy
eigenstates of the Hamiltonian, which should therefore be
diagonalized in terms of certain creation and annihilation
operators.
\subsection{Calculation of the free Hamiltonian}
We shall neglect that part of the Hamiltonian depending on the
Faddeev-Popov ghost fields $\oc^a$ and $c^a$, as well as depending
on the fields $\alpha^{i,a}$, $\overline{\al}^{i,a}$, $
\beta^{i,a}$, $ \overline{\beta}^{i,a}$, $\chi^{+,i,a}$ and
$\chi^{-,i,a}$, as these are of no relevance for the present
discussion, and we call this ``reduced'' Hamiltonian
$\mathcal{H}_0^\prime$. Since
\begin{equation}\label{ham1}
    \mathcal{H}_0^\prime= \int d^3x\left[ \pi_i \p_0 A^i+b\p_0A_0+\pi_{N,i}\p_0 N^i+\pi_{L,i}\p_0 L^i-\mathcal{L}\right]\,,
\end{equation}
we find
\begin{eqnarray}\label{ham2}
    \mathcal{H}_0^\prime&=& \int d^3x\left[
-b\p_\mu
A^\mu-\frac{1}{2}b^2+(F^{0i}+mN^i)\p_0A^i+b\p_0A_0+(\p_0N^i-mA^i)\p_0N^i\right.\nonumber\\
&-&\left.\p_0L^i\p_0L^i-\frac{1}{2}F^{0i}F^{0i}+\frac{1}{4}F^{ij}F^{ij}-mN^i(\p^0A^i-\p^iA^0)+m\varepsilon^{ijk}L^k\p^iA^j\right.\nonumber\\
&-&\left.\frac{1}{2}\p_0N^i\p_0N^i+\frac{1}{2}N^i\p_j\p^jN^i+\frac{1}{2}\p_0L^i\p_0L^i-\frac{1}{2}L^i\p_j\p^jL^i-\frac{1}{2}\wm^2L^iL^i+\frac{1}{2}\wm^2N^iN^i\right]\,.\nonumber\\
\end{eqnarray}
Since we must plug in the solutions (\ref{mu11}) into the field
expression of the Hamiltonian $\mathcal{H}_0^\prime$ to find its
operator valued expression, we can already use the equations of
motions (\ref{z1a})-(\ref{z1d}) to simplify a bit
$\mathcal{H}_0^\prime$. Since
\begin{eqnarray}\label{ham3}
  F^{0i}\equiv\p^0A^i-\p^i A^0 &=& \frac{1}{m}\left(\p^2+\wm^2\right)N^i \,,\nonumber\\
&\mbox{and}&\nonumber\\
  \varepsilon^{ijk}\p^iA^j &=&
\frac{1}{m}\left(\p^2+\wm^2\right)L^k\nonumber\\ \Rightarrow
F^{pq}\equiv\p^pA^q-\p^qA^p &=&
\frac{1}{m}\varepsilon^{ipq}\left(\p^2+\wm^2\right)L^i \nonumber\\
\Rightarrow \left(\p^pA^q-\p^qA^p\right)^2 &=&
\frac{2}{m^2}\left(\p^2+\wm^2\right)L^i\left(\p^2+\wm^2\right)L^i\,,
\end{eqnarray}
we simplify $\mathcal{H}_0^\prime$ to
\begin{eqnarray}\label{ham4}
    &&\left.\mathcal{H}_0^\prime\right|_{\mbox{\tiny on-shell}}\nonumber\\&=& \int d^3x\left[
\frac{1}{2}b^2+\frac{1}{m}(\p^2+M^2)N^i\p_0A^i+b\p_0A_0+(\p_0N^i-mA^i)\p_0N^i\right.\nonumber\\
&-&\left.\p_0L^i\p_0L^i-\frac{1}{2m^2}(\p^2+\wm^2)N^i(\p^2+\wm^2)N^i+\frac{1}{2m^2}(\p^2+\wm^2)L^i(\p^2+\wm^2)L^i\right.\nonumber\\&-&\left.N^i(\p^2+\wm^2)N^i+L^i(\p^2+\wm^2)L^i\right.\nonumber\\
&-&\left.\frac{1}{2}\p_0N^i\p_0N^i+\frac{1}{2}N^i\p_j\p^jN^i+\frac{1}{2}\p_0L^i\p_0L^i-\frac{1}{2}L^i\p_j\p^jL^i-\frac{1}{2}\wm^2L^iL^i+\frac{1}{2}\wm^2N^iN^i\right]\,.\nonumber\\
\end{eqnarray}
As a first exercise, let us have a look at the terms that give rise
to the $\widetilde{\ell}_3$-oscillator. Due to (\ref{mu16}), it is
easy to see that the only terms relevant for this are
\begin{eqnarray}\label{ham4bis}
   &&\int d^3x\left[
-\p_0L^i\p_0L^i+\frac{1}{2}\p_0L^i\p_0L^i-\frac{1}{2}L^i\p_j\p^jL^i-\frac{1}{2}\wm^2L^iL^i\right]\nonumber\\
&=&-2\int\frac{d^3q}{(2\pi)^3}(\vq^2+\wm^2)\widetilde{\ell}_3^\dagger(\vq)\widetilde{\ell}_3(\vq)\mbox{
after normal ordering}\,.
\end{eqnarray}
We immediately notice that we have a negative sign in the
Hamiltonian as far as the $\widetilde{\ell}_3$-modes are concerned,
which is inconsistent with a positive commutator for the reason of
positivity (see also section 7). Invoking a negative commutator then
necessarily leads to negative norm states. It might be important to
notice here that the BRST charge (\ref{z18}) is of no help
whatsoever to eliminate the negative norm states created by the
oscillator $\widetilde{\ell}_3^\dagger$. The charge $\mathcal{Q}_0$
does only act on massless states, as implied by (\ref{z18}) and
(\ref{mu30}).

\noindent One might wonder if there might exist a way out, in order
to find a positive Hamiltonian, without the need for negative norm
states. If we could change the sign of the ``new'' terms in the
gauge model, we should at least be able to avoid the problem in the
$\widetilde{\ell}_3$ sector.

\noindent We would thus like to use the following action
\begin{eqnarray}
  S_{phys}^\prime &=& S_{cl} +S_{gf}\;,\label{completeactionacc}\\
  S_{cl}^\prime&=&\int d^4x\left[-\frac{1}{4}F_{\mu \nu }^{a}F_{\mu \nu }^{a}-\frac{im}{4}(B-\overline{B})_{\mu\nu}^aF_{\mu\nu}^a
  -\frac{1}{4}\left( \overline{B}_{\mu \nu
}^{a}D_{\sigma }^{ab}D_{\sigma }^{bc}B_{\mu \nu
}^{c}-\overline{G}_{\mu \nu }^{a}D_{\sigma }^{ab}D_{\sigma
}^{bc}G_{\mu \nu
}^{c}\right)\right.\nonumber\\
&+&\left.\frac{3}{8}%
m^{2}\lambda _{1}\left( \overline{B}_{\mu \nu }^{a}B_{\mu \nu
}^{a}-\overline{G}_{\mu \nu }^{a}G_{\mu \nu }^{a}\right)
-m^{2}\frac{\lambda _{3}}{32}\left( \overline{B}_{\mu \nu
}^{a}-B_{\mu \nu }^{a}\right) ^{2}\right.\nonumber\\&+&\left.
\frac{\lambda^{abcd}}{16}\left( \overline{B}_{\mu\nu}^{a}B_{\mu\nu}^{b}-\overline{G}_{\mu\nu}^{a}G_{\mu\nu}^{b}%
\right)\left( \overline{B}_{\rho\sigma}^{c}B_{\rho\sigma}^{d}-\overline{G}_{\rho\sigma}^{c}G_{\rho\sigma}^{d}%
\right) \right]\;, \label{completeactionbacc}
\end{eqnarray}
which is completely similar to the action \eqref{completeaction} up
to a few signs.

 \noindent The steps in the algebraic
renormalizability analysis of \cite{Capri:2005dy,Capri:2006ne} are
not essentially affected by these sign changes. More precisely, it
is still possible to show the renormalizability to all orders. It is
clear that the supersymmetry (\ref{supersym}) is still present for
$m=0$, so that the equivalence with ordinary Yang-Mills theories is
maintained for $m\equiv0$.

\noindent In the decomposed form, we have the following (quadratic)
action in Minkowski space time
\begin{eqnarray}\label{last5bnieuw}
  S_0&=&\int d^4x\left[-\frac{1}{4}\left(\p_\mu A_\nu^a-\p_\nu A_\mu^a\right)^2-mN^{i,a}\left(\p^0 A^{i a}-\p^i A^{0 a}\right)
+m\varepsilon^{ijk}L^{k,a}\left(\p^i A^{j a}\right)
  \right.\nonumber\\&-&\left.\frac{1}{2} \chi^{+,i
,a}\p^2 \chi^{-, i, a}+\frac{1}{2} N^{i ,a}\p^2 N^{i, a}-\frac{1}{2}
L^{i ,a}\p^2 L^{i,
a}-\frac{1}{2}\overline{\beta}^{i,a}\p^2 \beta^{i,a}+\frac{1}{2}\overline{\al}^{i,a}\p^2 \al^{i,a}\right.\nonumber\\&+&\left.\frac{3}{4}%
m^{2}\lambda _{1}\left(
\chi^{+,i,a}\chi^{-,i,a}-N^{i,a}N^{i,a}+L^{i,a}L^{i,a}+\overline{\beta}^{i,a}
\beta^{i,a}-\overline{\al}^{i,a} \al^{i,a}\right)
\right.\nonumber\\&-&\left. m^{2}\frac{\lambda
_{3}}{4}\left(N^{i,a}N^{i,a}-L^{i,a}L^{i,a}\right)+b^{a}%
\partial _{\mu }A^{\mu a}+\overline{c}^{a}\p^2
c^{a}+\frac{\xi }{2}b^{a}b^{a}\right]\,.
\end{eqnarray}

\noindent Evidently, the cohomology analysis of section 4 can be
immediately translated into the new language.

\noindent The classical equations of motion read
\begin{eqnarray}
\label{z1anieuw}b+\p^\mu A_\mu&=&0\,,\\
     \label{z1bnieuw}\p^2A_0+m\p_i N^{i}&=&0\,,\\
\label{z1cnieuw}(\p^2+\widetilde{m}^2)N^{i}-m(\p^0
A^{i}-\p^iA^{0})&=&0\,, \\
\label{z1enieuw}-(\p^2+\widetilde{m}^2)L^{i}-m\varepsilon^{ijk}\p_j
A^k&=&0\,,\\
\label{z1dnieuw}\p^2 A^{i}-m\p_0 N^{i}+m\varepsilon^{ijk}\p_j
L^{k}&=&0\,.
\end{eqnarray}
We solve them by using once more the Fourier decomposition
(\ref{mu11}), yielding the following differential equations in time
$t$
\begin{eqnarray}\label{nieuw1}
(\p_0^2+\vq^2)\widehat{A}_0&=&-mi|\vec{q}|\widehat{N}_3\,,\nonumber\\
(\p_0^2+\vq^2+\widetilde{m}^2)\widehat{N}_3&=&m\left(\p_0 \widehat{A}_3+i|\vec{q}|\widehat{A}_0\right)\,,\nonumber\\
(\p_0^2+q^2)\widehat{A}_3-m\p_0\widehat{N}_3&=&0\,,
\end{eqnarray}
and
\begin{eqnarray}\label{nieuw2}
(\p_0^2+\vq^2+\widetilde{m}^2)\widehat{N}_{1,2}&=&m\p_0
\widehat{A}_{1,2}\,,\nonumber\\
(\p_0^2+q^2+\widetilde{m}^2)\widehat{L}_3&=&0\,,\nonumber\\
(\p_0^2+q^2+\widetilde{m}^2)\widehat{L}_1&=&im|\vec{q}|\widehat{A}_2\,,\nonumber\\
(\p_0^2+q^2+\widetilde{m}^2)\widehat{L}_2&=&-im|\vec{q}|\widehat{A}_1\,,\nonumber\\
(\p_0^2+q^2)\widehat{A}_1-m\p_0
\widehat{N}_1&=&im|\vec{q}|\widehat{L}_2\,,\nonumber\\(\p_0^2+q^2)\widehat{A}_2-m\p_0
\widehat{N}_2&=&-im|\vec{q}|\widehat{L}_1\,.
\end{eqnarray}
Decoupling the first set of equations (\ref{nieuw1}) leads to
\begin{eqnarray}\label{nieuw3}
    (\p_0^2+\vq^2)^2(\p_0^2+\vec{q}^2+M^2)\widehat{A}_{0,3}&=&0\,,\nonumber\\
(\p_0^2+\vq^2)(\p_0^2+\vec{q}^2+M^2)\widehat{N}_{3}&=&0\,,
\end{eqnarray}
where the mass $M^2$ is now defined by
\begin{equation}\label{nieuw4}
    M^2=\wm^2-m^2\,.
\end{equation}
The presence of the mass scale $\wm^2\neq0$ is essential here to
avoid the appearance of a tachyonic mass $-m^2$, which is due to the
change of sign in the action (\ref{completeactionacc}). The mass
scales $\propto \lambda_im^2$ leading to the mass $\wm^2$ were
originally introduced in \cite{Capri:2005dy} to ensure that the
final action is renormalizable from the algebraic point of view. In
the present context, these also play a physical role in order to
avoid tachyons, if we assume that $\wm^2>m^2$.

\noindent The solution to (\ref{nieuw3}) is still given by
(\ref{mu20}), however the relations between the coefficients are
modified into
\begin{eqnarray}\label{mu21bis}
  M^2\widehat{a}_0&=& mi|\vec{q}|\widehat{n}_3\,, \nonumber\\
   2\widetilde{a}_0&=&mn_3\,,\nonumber\\
    \widetilde{a}_3 &=& \widetilde{a}_0\,,\nonumber\\
-mi|\vq|a_3+m\widetilde{a}_3+mi|\vq|a_0&=&\wm^2n_3\,, \nonumber\\
M^2\widehat{a}_3&=&mi\sqrt{M^2+\vq^2}\widehat{n}_3\,.
\end{eqnarray}
It is straightforward to verify that
\begin{eqnarray}\label{nieuwbveld}
    b(x)&=&\int\frac{d^3q}{(2\pi)^3}
\left(\frac{M^2}{m}\right)n_3(\vq)e^{-i|\vq|t}e^{i\vq\vec{x}}+\mbox{h.c.}\,.
\end{eqnarray}
The conjugate momenta are given by
\begin{eqnarray}\label{z14}
  \pi_{i}^a &=& \frac{\delta\cal L}{\delta (\p_0 A^{i,a})}=-F_{0i}^a+mN_i^a\,,\nonumber \\
\pi_{N,i}^a&=&\frac{\delta\cal L}{\delta (\p_0 N^{i,a})}=\p_{0}N_{i}^a-mA_i^a\,,\nonumber \\
\pi_{L,i}^a&=&\frac{\delta\cal L}{\delta (\p_0 L^{i,a})}=-\p_0 L_i^a\,,\nonumber\\
  \pi^{0,a} &=& b^a\equiv -\p^{\mu} A_{\mu}^a\,.
\end{eqnarray}
The equation of motion for $\widehat{L}_3$ is immediately solved by
(\ref{mu22}).

\noindent Let us now turn to the $(1,2)$-sector. It is not difficult
to check that (\ref{mu23}) and (\ref{mu24}) still holds, but now
with
\begin{eqnarray}\label{qqrel}
\begin{array}{lll}
  \wm^2 n_{1} = -im|\vq| a_1\,, & \quad & \wm^2 n_{2} = -im|\vq| a_2\,,\\
  m \widehat{n}_{1} = -i\sqrt{\vq^2+M^2} \widehat{a}_1 \,, & \quad& m \widehat{n}_{2} = -i\sqrt{\vq^2+M^2} \widehat{a}_2\,, \\
   m \widehat{\ell}_1= i|\vq|\widehat{a}_2 \,, & \quad& m \widehat{\ell}_2= -i|\vq|\widehat{a}_1\,, \\
  n_1=\ell_2 \,, & \quad& n_2=-\ell_1\,,\\
  \sqrt{\vq^2+\wm^2}\widetilde{n}_1=|\vq|\widetilde{\ell}_2 \,, &
\quad&\sqrt{q^2+\wm^2}\widetilde{n}_2=-|\vq|\widetilde{\ell}_1\,.
\end{array}
\end{eqnarray}
The (reduced) free Hamiltonian $\mathcal{H}_0^\prime$ becomes
\begin{eqnarray}\label{ham2bis}
    \mathcal{H}_0^{\prime}&=& \int d^3x\left[
-b\p_\mu
A^\mu-\frac{1}{2}b^2+(F^{0i}-mN^i)\p_0A^i+b\p_0A_0+mA^i\p_0N^i\right.\nonumber\\
&-&\left.\frac{1}{2}F^{0i}F^{0i}+\frac{1}{4}F^{ij}F^{ij}+mN^i(\p^0A^i-\p^iA^0)-m\varepsilon^{ijk}L^k\p^iA^j\right.\nonumber\\
&-&\left.\frac{1}{2}\p_0N^i\p_0N^i-\frac{1}{2}N^i\p_j\p^jN^i+\frac{1}{2}\p_0L^i\p_0L^i+\frac{1}{2}L^i\p_j\p^jL^i+\frac{1}{2}\wm^2L^iL^i-\frac{1}{2}\wm^2N^iN^i\right]\,.\nonumber\\
\end{eqnarray}
Since the relations (\ref{ham3}) are still valid, we find for the
first part
\begin{eqnarray}\label{ham4bis}
    &&\left.\mathcal{H}_0^{\prime}\right|_{\mbox{\tiny on-shell}}\nonumber\\&=& \int d^3x\left[
\frac{1}{2}b^2+\frac{1}{m}(\p^2+M^2)N^i\p_0A^i+b\p_0A_0+mA^i\p_0N^i\right.\nonumber\\
&-&\left.\frac{1}{2m^2}(\p^2+\wm^2)N^i(\p^2+\wm^2)N^i+\frac{1}{2m^2}(\p^2+\wm^2)L^i(\p^2+\wm^2)L^i\right.\nonumber\\&+&\left.N^i(\p^2+\wm^2)N^i-L^i(\p^2+\wm^2)L^i-\frac{1}{2}\p_0N^i\p_0N^i-\frac{1}{2}N^i\p_j\p^jN^i\right.\nonumber\\
&+&\left.\frac{1}{2}\p_0L^i\p_0L^i+\frac{1}{2}L^i\p_j\p^jL^i+\frac{1}{2}\wm^2L^iL^i-\frac{1}{2}\wm^2N^iN^i\right]\,.
\end{eqnarray}
After some tedious algebra, we can express the part of the
Hamiltonian corresponding to the (massive) $\sim$-modes as
\begin{eqnarray}\label{ham5}
    \mathcal{H}_0^{\prime,(1)}&=&\int \frac{d^3q}{(2\pi)^3}\left[2\left(\vq^2+\wm^2\right)\widetilde{\ell}_3^\dagger(\vq)\widetilde{\ell}_3(\vq)+2\left(\vq^2+\wm^2\right)\widetilde{\ell}_1^\dagger(\vq)\widetilde{\ell}_1(\vq)+
2\left(\vq^2+\wm^2\right)\widetilde{\ell}_2^\dagger(\vq)\widetilde{\ell}_2(\vq)\right.\nonumber\\&-&\left.2\left(\vq^2+\wm^2\right)\widetilde{n}_1^\dagger(\vq)\widetilde{n}_1(\vq)-2\left(\vq^2+\wm^2\right)\widetilde{n}_2^\dagger(\vq)\widetilde{n}_2(\vq)\right]\,.
\end{eqnarray}
Using the relations (\ref{mu21bis}) and (\ref{qqrel}), we rewrite
(\ref{ham5}) as
\begin{eqnarray}\label{ham6}
    \mathcal{H}_0^{\prime,(1)}&=&\int \frac{d^3q}{(2\pi)^3}\left[2\left(\vq^2+\wm^2\right)\widetilde{\ell}_3^\dagger(\vq)\widetilde{\ell}_3(\vq)+2\wm^2\widetilde{\ell}_1^\dagger(\vq)\widetilde{\ell}_1(\vq)+
2\wm^2\widetilde{\ell}_2^\dagger(\vq)\widetilde{\ell}_2(\vq)\right]\,.
\end{eqnarray}
The second part, corresponding the (massive) $\wedge$-modes is given
by
\begin{eqnarray}\label{ham7}
    \mathcal{H}_0^{\prime,(2)}&=&\int \frac{d^3q}{(2\pi)^3}\left[-2(\vq^2+M^2)\left(1+\frac{m^2}{M^2}\right)\widehat{n}_3^\dagger(\vq)\widehat{n}_3(\vq)
-2\wm^2\widehat{n}_1^\dagger(\vq)\widehat{n}_1(\vq)-2\wm^2\widehat{n}_2^\dagger(\vq)\widehat{n}_2(\vq)\right]\,.\nonumber\\
\end{eqnarray}
Finally, we also have a part $\mathcal{H}_0^{\prime,(3)}$
corresponding to the massless modes, which we do not write in more
detail.

\sect{Fixing the commutation relations between creation and
annihilation operators and the non-unitarity of the model}
\subsection{Outline of the idea}
We recall here that the equations of motion
(\ref{z1anieuw})-(\ref{z1dnieuw}) constitute several (nontrivial)
relations between the field components. Unfortunately, these
relations are quite complicated to solve explicitly. Closely related
to this is the fact that it is not immediately clear how to
diagonalize the quadratic form appearing in the action
(\ref{last5bnieuw}). Moreover, this would-be diagonalized form
should also be of maximum second order in the derivatives to allow
for a consistent quantization.

\noindent Since the fields are related to each other, we can expect
the same for the conjugate momenta. E.g., it is easily checked that
$\pi_0(\vec{x},t)=-\pi_3(\vec{x},t)$.

\noindent It would appear that quantization, starting from
configuration space, is a highly nontrivial task due to the existing
relationships between the fields and momenta. It would be beneficial
to quantize the theory directly in Fourier (momentum) space, in
which case the independent degrees of freedom are clearly
identifiable.

\noindent There does exist a simple way to derive the appropriate
(anti-)commutation relations between the creation and annihilation
operators, when we rely on the Heisenberg picture. We can always
calculate the free Hamiltonian operator $\mathcal{H}_0$ without
exact knowledge of the commutation relations. The precise value of
the commutators only influences the constant part of
$\mathcal{H}_0$, which we can drop after normal ordering. On one
hand, we already know the time evolution of all fields $\Psi$, as
written down in (\ref{mu11}) by using the Fourier decomposition and
the solutions (\ref{mu20}) and (\ref{mu24}). On the other hand,
these fields are treated as operators in the Heisenberg picture, and
consequently their time evolution is dictated by the Heisenberg
equation
\begin{equation}\label{heis1}
    \left[\mathcal{H}_0,\Psi\right]=-i\frac{\p}{\p t}\Psi\,.
\end{equation}
Demanding consistency allows to fix the commutators between the
creation and annihilation operators, which are supposed to be number
valued. As an illustration, consider the Klein-Gordon action
\begin{equation}\label{KG1}
    S_{KG}=\int d^4x\left(
-\frac{1}{2}\varphi\left(\p^2+m^2\right)\varphi\right)\,.
\end{equation}
The solution to the equation of motion is expressed by
\begin{eqnarray}\label{KG2}
    \varphi(\vec{x},t)&=&\int\frac{d^3q}{(2\pi)^3}\frac{1}{2\omega_q}a(\vq)e^{-i\omega_qt}e^{i\vec{q}\vec{x}}+\mbox{h.c.}\nonumber\\
\omega_q&=&\sqrt{\vq^2+m^2}\,.
\end{eqnarray}
As it is well known, the normal ordered Hamiltonian is given
by\footnote{We disregard the conventional $:\mathcal{H}_{KG}:$
notation and continue to use $\mathcal{H}_{KG}$.}
\begin{eqnarray}\label{KG3}
    \mathcal{H}_{KG}&=&\int\frac{d^3q}{(2\pi)^3}\frac{1}{2\omega_q} \omega_q a^\dagger(\vq)
a(\vq)\,,
\end{eqnarray}
so that requiring
\begin{equation}\label{KG4}
    \left[\mathcal{H}_{KG},\varphi(\vec{x},t)\right]=-i\frac{\p}{\p
t}\varphi(\vec{x},t)\,,
\end{equation}
leads to the correct commutation relation
\begin{equation}\label{KG5}
    \left[a(\vq),a^\dagger(\vk)\right]=2\omega_q(2\pi)^3\delta^{(3)}(\vq-\vk)\,.
\end{equation}
\subsection{Quantization of the massive gauge model using the Hamiltonian $\mathcal{H}_0^\prime$}
In order to find the commutators for the massive $\sim$-modes, we
only need to consider $\mathcal{H}_0^{\prime,(1)}$. When we impose
\begin{eqnarray}\label{app1}
    \left[\mathcal{H}_0^{\prime,(1)},\widehat{L}^i(\vq,t)\right]=-i\frac{\p}{\p
t}\widehat{L}^i(\vq,t)\,,
\end{eqnarray}
we deduce that the following commutation relations arise
\begin{eqnarray}
  \left[\widehat{\ell}_{1,2}(\vq),\widehat{\ell}_{1,2}^\dagger(\vk)\right] &=& (2\pi)^3\frac{\sqrt{\vq^2+\wm^2}}{2\wm^2}\delta^{(3)}(\vk-\vq)\,, \nonumber\\
    \left[\widehat{\ell}_{3}(\vq),\widehat{\ell}_{3}^\dagger(\vk)\right] &=& (2\pi)^3\frac{1}{2\sqrt{\vq^2+\wm^2}}\delta^{(3)}(\vk-\vq)\,. \nonumber\\
\mbox{others trivial}&&
\end{eqnarray}
We can define novel operators as follows
\begin{eqnarray}\label{heis2}
    \widetilde{\ell}_{1,2}(\vq)&=&\frac{1}{2\wm}\widetilde{\ell}_{1,2}^\prime(\vq)\,,\nonumber\\
    \widetilde{\ell}_{3}(\vq)&=&\frac{1}{2\sqrt{\vq^2+\wm^2}}\widetilde{\ell}_{3}^\prime(\vq)\,,
\end{eqnarray}
to arrive at a free Hamiltonian in a more standard form
\begin{eqnarray}\label{heis5}
    \mathcal{H}_0^{\prime,(1)}&=&\int \frac{d^3q}{(2\pi)^3}\frac{1}{2\widetilde{\omega}_q}\left[2\widetilde{\omega}_q\sum_{i=1}^3\widetilde{\ell}_{i}^{\prime\dagger}(\vq)\widetilde{\ell}_i^\prime(\vq)
\right]\,,
\end{eqnarray}
with commutation relations
\begin{eqnarray}\label{heis6}
  \left[\widetilde{\ell}_{i}^\prime(\vq),\widetilde{\ell}_{j}^{\prime\dagger}(\vk)\right] &=& \widetilde{\omega}_q(2\pi)^3\delta_{ij}\delta^{(3)}(\vk-\vq)\,, \nonumber\\
\mbox{others trivial\,.}&&
\end{eqnarray}
The energy was defined as
\begin{eqnarray}\label{heis7}
 \widetilde{\omega}_q &=& \sqrt{\vq^2+\wm^2}\,.
\end{eqnarray}
We conclude that the modes $\widetilde{\ell}_{i}^\prime(\vq)$
correspond to the 3 polarizations of a vector particle with mass
$\wm$. As it is clear from (\ref{heis6}), they have a positive norm.

\noindent If we would have used the original Hamiltonian
(\ref{ham4}), we would have found a negative sign commutator, as
already advocated below (\ref{ham4bis}).

\noindent In a completely similar fashion, we can quantize the
$\wedge$-modes. We rewrite (\ref{ham7}) into the form
\begin{eqnarray}\label{heis5}
    \mathcal{H}_0^{\prime,(2)}&=&\int \frac{d^3q}{(2\pi)^3}\frac{1}{2\widehat{\omega}_q}\left[-\widehat{\omega}_q\sum_{i=1}^3\widehat{n}_{i}^{\prime\dagger}(\vq)\widehat{n}_i^\prime(\vq)
\right]\,,
\end{eqnarray}
by a suitable rescaling of the fields, in which case the
corresponding commutation relations become
\begin{eqnarray}\label{heis9}
  \left[\widehat{n}_{i}^\prime(\vq),\widehat{n}_{j}^{\prime\dagger}(\vk)\right] &=& -2\widehat{\omega}_q(2\pi)^3\delta_{ij}\delta^{(3)}(\vk-\vq) \nonumber\\
\mbox{others trivial\,.}&&
\end{eqnarray}
with
\begin{eqnarray}\label{heis7}
 \widehat{\omega}_q &=& \sqrt{\vq^2+M^2}\,.
\end{eqnarray}
We conclude that we have been able to change the negative norms
corresponding to the $\widetilde{\ell}_i$-modes in our first model
(\ref{completeaction}) by passing to (\ref{completeactionbacc}), but
unfortunately the negative norm problem has been ``shifted'' to the
$\widehat{n}_i$-modes, as it is clearly visible from (\ref{heis9}).
We are thus forced to conclude that our theory is not unitary, as we
do not seem to be able to remove these negative norm states,
corresponding to the polarizations of a massive vector particle with
mass $M$. Just as in the case of the $\widetilde{\ell}_i$-modes, we
reckon once more that the BRST symmetry has nothing to say about the
$\widehat{n}_i$-modes.

\noindent To completely finish the analysis, we should in principle
also determine the free Hamiltonian corresponding to the massless
modes. However, this has become a bit redundant job by now since we
already have established that the theory cannot be unitary. Quite
obviously, some restriction on the allowed (massless) states and
Faddeev-Popov ghost states will arise from the BRST charge
(\ref{z18}). However, this might also turn out to be a not so
trivial task after all due to the multipole character, encoded in
(\ref{mu20}) in the terms linear in time $t$. As it was found in the
case of the toy model of \cite{Ferrara:1977mv}, the Hamiltonian is
not even necessarily diagonal, obscuring the particle
interpretation. It is even noticed that the Hamiltonian cannot be
diagonalized with linear transformations. As already said, we shall
however not dwell on that point here.

\sect{Conclusion} We continued our investigation of a recently
proposed local non-Abelian gauge invariant action containing mass
terms (\ref{completeaction}), which can be renormalized to all
orders of perturbation once a gauge is chosen. Despite the fact that
the model enjoys a BRST symmetry with nilpotent generator given in
(\ref{brst3}), it turned out to be impossible to remove all negative
norm states from the physical state space. As a consequence, our
model is not unitary and thus not useful as a physical theory
containing massive particles as asymtotic observables.

\noindent As already indicated in the introduction, this does not
mean that our action is now useless. We can still wonder whether it
could be generated dynamically in a nonperturbative fashion when we
would start with the initially massless version of our gauge model.
As a matter of fact, we can propose our model as an alternative to
ordinary Yang-Mills theory in the perturbative (massless) region,
since both are equivalent \cite{Capri:2006ne}. It might however be
more convenient to search for a dynamical mass $m$ in our model
since the mass term can be at least coupled to the action without
spoiling the renormalizability or gauge invariance. Of course, it
remains to be investigated if a sensible gap equation could be
established. If this would work out, we might have a gauge invariant
mechanism behind the dynamical generation of a massive parameter
into a gauge theory.

\noindent We repeat that the generation of a mass would not be
necessarily in conflict with the nonunitarity of the massive gauge
model, since we start from the (unitary) massless theory and we do
no longer want to describe the asymptotic high energy behaviour
wherefore perturbation theory applies perfectly, but rather we are
entering a phenomenologically interesting region where e.g. the
gluons already loose their physical meaning as an observable.

\noindent Since our generated mass would be gauge invariant, it
could enter physical correlation functions. As such, it could be
investigated if it could serve as a possible alternative to the
$\langle A^2\rangle_{\min}$ condensate used to explain the
$\frac{1}{Q^2}$ power corrections
\cite{Gubarev:2000eu,Gubarev:2000nz}.

\noindent We conclude by noticing that we could repeat our analysis
in the Abelian case. Then it can be shown that the action
(\ref{completeaction}) is stable without the need for $\lambda_1$
and $\lambda_3$ \cite{Capri:2005dy}. It can also be shown that the
model is equivalent with the Abelian Stueckelberg model
\cite{Ruegg:2003ps}, described by
\begin{eqnarray}
  S_{\mathrm{stuck}} &=& -\frac{1}{4}\int d^4x F_{\mu\nu} F^{\mu\nu}+\frac{m^2}{2}\int d^4x (A_\mu+\p_\mu \phi)(A^\mu+\p^\mu \phi)\nonumber\\&+&\int d^4x\left(b\p A+\frac{1}{2}b^2+\oc \p^2
c\right)\,.
\end{eqnarray}
When the auxiliary fields are integrated out in both cases, some
manipulation leads to the same (nonlocal) action
\begin{eqnarray}\label{actie}
  S_{\mathrm{nonlocal}} &=& -\frac{1}{4}\int d^4x F_{\mu\nu} F^{\mu\nu}-\frac{m^2}{4}\int d^4x F_{\mu\nu}\frac{1}{\p^2}F^{\mu\nu}+\int d^4x\left(b\p A+\frac{1}{2}b^2+\oc \p^2
c\right)
\end{eqnarray}
It is known that the Abelian Stueckelberg model is renormalizable
and unitary, see e.g. \cite{Ruegg:2003ps} for a review. If we would
analyze the unitarity of the Abelian version of
(\ref{completeactionb}), we would run into exactly the same problem
as in the non-Abelian case, i.e. the presence of negative norm
states in the physical subspace.

\noindent The lesson to be learnt is the following. We depart from
the same nonlocal action, but in order to give a consistent
quantization of the theory, including an analysis of the unitarity
and renormalizability, we are forced to bring the action in some
localized polynomial form. Apparently, the exact procedure of
localization affects these results. For example, in the Abelian
case, there are 2 quite distinct approaches to bring the action
(\ref{actie}) in a local and in addition renormalizable form: the
Stueckelberg approach or the pathway we followed. However, only the
Stueckelberg way gives a unitary model. Apparently, the precise role
of the additional fields that are introduced cannot be
underestimated in the discussion of the unitarity and/or
renormalizability.

\section*{Acknowledgments}
D.~Dudal would like to thank S.~P.~Sorella for useful discussions.
D.~Dudal is a postdoctoral fellow of the \emph{Special Research
Fund} of Ghent University

\section*{Appendix} Since we have been working with the decomposed
fields, the covariance of the (naive) commutation relations
(\ref{q1}) is obscured. It is worth having a look at this. We
decompose $B_{\mu\nu}$ and its complex conjugate $\oB_{\mu\nu}$ into
their real and imaginary part.
\begin{eqnarray}
   X_{\mu\nu}&=&\frac{1}{2}(B_{\mu\nu}+\overline{B}_{\mu\nu})\qquad\qquad   Y_{\mu\nu}=\frac{1}{2i}(B_{\mu\nu}-\overline{B}_{\mu\nu})\,,
\end{eqnarray}
then
\begin{eqnarray}
\begin{array}{lll}
  M_i = X_{0i} \,,& \quad&N_i = Y_{0i}\,, \\
  K^i =\frac{1}{2}\varepsilon^{ijk}X_{jk} \,,&
\quad&L^i=\frac{1}{2}\varepsilon^{ijk}Y_{jk}\,.
\end{array}
\end{eqnarray}
Covariance and the antisymmetry would require commutation relations
like
\begin{eqnarray}
   \left[X_{\mu\nu}(\vec{x},t),\pi_{\alpha\beta}^X(\vec{y},t)\right]&=&i \left(g_{\mu\alpha}g_{\nu\beta}-g_{\mu\beta}g_{\nu\alpha}\right)\delta^{(3)}(\vec{x}-\vec{y})\,,\nonumber\\
   \left[Y_{\mu\nu}(\vec{x},t),\pi_{\alpha\beta}^Y(\vec{y},t)\right]&=&i
\left(g_{\mu\alpha}g_{\nu\beta}-g_{\mu\beta}g_{\nu\alpha}\right)\delta^{(3)}(\vec{x}-\vec{y})\,.
\end{eqnarray}
Specifically
\begin{eqnarray}\label{app1}
   \left[X_{0i}(\vec{x},t),\pi_{0j}^X(\vec{y},t)\right]&=&i g_{ij}\delta^{(3)}(\vec{x}-\vec{y})=-i\delta_{ij}\delta^{(3)}(\vec{x}-\vec{y})\,,\nonumber\\
   \left[X_{ij}(\vec{x},t),\pi_{kl}^X(\vec{y},t)\right]&=&i
\left(\delta_{ik}\delta_{jl}-\delta_{il}\delta_{jk}\right)\delta^{(3)}(\vec{x}-\vec{y})\,,
\end{eqnarray}
and the same for $Y$. Consequently
\begin{eqnarray}
   \left[M_{i}(\vec{x},t),\pi_{j}^M(\vec{y},t)\right]&=&-i\delta_{ij}\delta^{(3)}(\vec{x}-\vec{y})\,,\nonumber\\
   \left[K_{r}(\vec{x},t),\pi_{s}^K(\vec{y},t)\right]&=&\frac{1}{4}\varepsilon^{rij}\varepsilon^{skl}\left[X_{ij}(\vec{x},t),\pi_{kl}^X(\vec{y},t)\right]\nonumber\\
&=&i\frac{1}{4}\varepsilon^{rij}\varepsilon^{skl}
\left(\delta_{ik}\delta_{jl}-\delta_{il}\delta_{jk}\right)\delta^{(3)}(\vec{x}-\vec{y})
=i\delta_{rs}\delta^{(3)}(\vec{x}-\vec{y})\,,
\end{eqnarray}
and the same for $N$ and $L$.


\begin{thebibliography}{99}
\bibitem{Capri:2005dy}
M.~A.~L.~Capri, D.~Dudal, J.~A.~Gracey, V.~E.~R.~Lemes,
R.~F.~Sobreiro, S.~P.~Sorella and H.~Verschelde, \emph{A study of
the gauge invariant, nonlocal mass operator $Tr \int d^4x F_{\mu\nu}
\frac{1}{D^2} F_{\mu\nu} $  in Yang-Mills theories},  Phys.\ Rev.\ D
{\bf 72} (2005) 105016.

\bibitem{Capri:2006ne}
  M.~A.~L.~Capri, D.~Dudal, J.~A.~Gracey, V.~E.~R.~Lemes, R.~F.~Sobreiro, S.~P.~Sorella and H.~Verschelde,
 \emph{Quantum properties of a non-Abelian gauge invariant action with a
mass parameter}, Phys.\ Rev.\ D {\bf 74} (2006) 045008.

\bibitem{vanRitbergen:1998pn}
 T.~van Ritbergen, A.~N.~Schellekens and J.~A.~M.~Vermaseren, \emph{Group theory factors for Feynman diagrams}, Int.\ J.\ Mod.\ Phys.\  A {\bf 14} (1999) 41.

\bibitem{Cornwall:1981zr}
J.~M.~Cornwall, \emph{Dynamical Mass Generation In Continuum QCD},
Phys.\ Rev.\ D {\bf 26} (1982) 1453.

\bibitem{Chetyrkin:1998yr}
K.~G.~Chetyrkin, S.~Narison and V.~I.~Zakharov, \emph{Short-distance
tachyonic gluon mass and $\frac{1}{Q^2}$ corrections}, Nucl.\ Phys.\
B {\bf 550} (1999) 353.

\bibitem{Gubarev:2000eu}
F.~V.~Gubarev, L.~Stodolsky and V.~I.~Zakharov, \emph{On the
significance of the vector potential squared}, Phys.\ Rev.\ Lett.\
{\bf 86} (2001) 2220.

\bibitem{Gubarev:2000nz}
F.~V.~Gubarev and V.~I.~Zakharov, \emph{On the emerging
phenomenology of $\left\langle A_\mu^2\right\rangle$}, Phys.\ Lett.\
B {\bf 501} (2001) 28.

\bibitem{Gribov:1977wm}
V.~N.~Gribov, \emph{Quantization Of Non-Abelian Gauge Theories},
Nucl.\ Phys.\ B {\bf 139} (1978) 1.

\bibitem{Semenov}  Semenov-Tyan-Shanskii and V.A. Franke, Zapiski Nauchnykh
Seminarov Leningradskogo Otdeleniya Matematicheskogo Instituta im.
V.A. Steklov AN SSSR, Vol. \textbf{120} (1982) 159. English
translation: New York: Plenum Press 1986.

\bibitem{Lavelle:1995ty}
 M.~Lavelle and D.~McMullan, \emph{Constituent quarks from QCD}, Phys.\ Rept.\  {\bf 279} (1997) 1.

\bibitem{Boucaud:2001st}
P.~Boucaud, A.~Le Yaouanc, J.~P.~Leroy, J.~Micheli, O.~Pene and
J.~Rodriguez-Quintero, \emph{Testing Landau gauge OPE on the lattice
with a $A^2$ condensate}, Phys.\ Rev.\ D {\bf 63} (2001) 114003.

\bibitem{Lavelle:1988eg}
M.~J.~Lavelle and M.~Schaden, \emph{Propagators And Condensates In
QCD}, Phys.\ Lett.\ B {\bf 208} (1988) 297.

\bibitem{Kondo:2001nq}
K.~I.~Kondo, \emph{Vacuum condensate of mass dimension 2 as the
origin of mass gap and quark confinement}, Phys.\ Lett.\ B {\bf 514}
(2001) 335.

\bibitem{Verschelde:2001ia}
H.~Verschelde, K.~Knecht, K.~Van Acoleyen and M.~Vanderkelen,
\emph{The non-perturbative groundstate of QCD and the local
composite operator $A_\mu^2$}, Phys.\ Lett.\ B {\bf 516} (2001) 307.

\bibitem{Dudal:2003vv}
D.~Dudal, H.~Verschelde, R.~E.~Browne and J.~A.~Gracey, \emph{A
determination of $\left\langle A_\mu^2\right\rangle$ and the
non-perturbative vacuum energy of Yang-Mills theory in the Landau
gauge}, Phys.\ Lett.\ B {\bf 562} (2003) 87.

\bibitem{Dudal:2002pq}
D.~Dudal, H.~Verschelde and S.~P.~Sorella, \emph{The anomalous
dimension of the composite operator $A^2$ in the Landau gauge},
Phys.\ Lett.\ B {\bf 555} (2003) 126.

\bibitem{Parisi:1980jy}
G.~Parisi and R.~Petronzio, \emph{On Low-Energy Tests Of QCD},
Phys.\ Lett.\  B {\bf 94} (1980) 51.

\bibitem{Halzen:1992vd}
F.~Halzen, G.~I.~Krein and A.~A.~Natale, \emph{Relating the QCD
pomeron to an effective gluon mass}, Phys.\ Rev.\ D {\bf 47} (1993)
295.

\bibitem{Field:2001iu}
J.~H.~Field, \emph{A phenomenological analysis of gluon mass effects
in inclusive radiative decays of the J/psi and Upsilon}, Phys.\
Rev.\ D {\bf 66} (2002) 013013.

\bibitem{Marenzoni:1994ap}
 P.~Marenzoni, G.~Martinelli and N.~Stella, \emph{The Gluon propagator on a large volume, at beta = 6.0},  Nucl.\ Phys.\  B {\bf 455} (1995)
339.

\bibitem{Langfeld:2001cz}
K.~Langfeld, H.~Reinhardt and J.~Gattnar, \emph{Gluon propagators
and quark confinement}, Nucl.\ Phys.\ B {\bf 621} (2002) 131.

\bibitem{Amemiya:1998jz}
K.~Amemiya and H.~Suganuma, \emph{Effective mass generation of
off-diagonal gluons as the origin of infrared Abelian dominance in
the maximally Abelian gauge in QCD}, Phys.\ Rev.\ D {\bf 60} (1999)
114509.

\bibitem{Bornyakov:2003ee}
V.~G.~Bornyakov, M.~N.~Chernodub, F.~V.~Gubarev, S.~M.~Morozov and
M.~I.~Polikarpov, \emph{Abelian dominance and gluon propagators in
the maximally Abelian gauge of SU(2) lattice gauge theory}, Phys.\
Lett.\ B {\bf 559} (2003) 214.

\bibitem{Aguilar:2004sw}
A.~C.~Aguilar and A.~A.~Natale, \emph{A dynamical gluon mass
solution in a coupled system of the Schwinger-Dyson
 equations}, JHEP {\bf 0408} (2004) 057.

\bibitem{Aguilar:2006gr}
A.~C.~Aguilar and J.~Papavassiliou, \emph{Gluon mass generation in
the PT-BFM scheme}, JHEP {\bf 0612} (2006) 012.

\bibitem{Dudal:2003gu}
D.~Dudal, H.~Verschelde, V.~E.~R.~Lemes, M.~S.~Sarandy,
S.~P.~Sorella and M.~Picariello, \emph{Gluon-ghost condensate of
mass dimension 2 in the Curci-Ferrari gauge}, Annals Phys.\  {\bf
308} (2003) 62.

\bibitem{Dudal:2003pe}
D.~Dudal, H.~Verschelde, V.~E.~R.~Lemes, M.~S.~Sarandy,
R.~F.~Sobreiro, S.~P.~Sorella, M.~Picariello and J.~A.~Gracey,
\emph{The anomalous dimension of the gluon-ghost mass operator in
Yang-Mills theory}, Phys.\ Lett.\ B {\bf 569} (2003) 57.

\bibitem{Dudal:2003np}
D.~Dudal, H.~Verschelde, V.~E.~R.~Lemes, M.~S.~Sarandy,
R.~F.~Sobreiro, S.~P.~Sorella and J.~A.~Gracey,
\emph{Renormalizability of the local composite operator $A_\mu^2$ in
linear covariant gauges}, Phys.\ Lett.\ B {\bf 574} (2003) 325.

\bibitem{Browne:2003uv}
 R.~E.~Browne and J.~A.~Gracey, \emph{Two loop effective potential for $\langle A_\mu^2\rangle$ in the Landau gauge in quantum
 chromodynamics}, JHEP {\bf 0311} (2003) 029.

\bibitem{Dudal:2003by}
D.~Dudal, H.~Verschelde, J.~A.~Gracey, V.~E.~R.~Lemes,
M.~S.~Sarandy, R.~F.~Sobreiro and S.~P.~Sorella, \emph{Dynamical
gluon mass generation from $\langle A_\mu^2\rangle$ in linear
covariant gauges}, JHEP {\bf 0401} (2004) 044.

\bibitem{Dudal:2004rx}
D.~Dudal, J.~A.~Gracey, V.~E.~R.~Lemes, M.~S.~Sarandy,
R.~F.~Sobreiro, S.~P.~Sorella and H.~Verschelde, \emph{An analytic
study of the off-diagonal mass generation for Yang-Mills theories in
the maximal Abelian gauge}, Phys.\ Rev.\ D {\bf 70} (2004) 114038.

\bibitem{Browne:2004mk}
R.~E.~Browne and J.~A.~Gracey, \emph{One loop MS-bar gluon pole mass
from the LCO formalism},  Phys.\ Lett.\  B {\bf 597} (2004) 368

\bibitem{Gracey:2004bk}
J.~A.~Gracey,  \emph{Two loop MS-bar gluon pole mass from the LCO
formalism}, Eur.\ Phys.\ J.\  C {\bf 39} (2005) 61.

\bibitem{Lemes:2006aw}
 V.~E.~R.~Lemes, R.~F.~Sobreiro and S.~P.~Sorella, \emph{Renormalizability of the dimension two gluon operator $A^2$ in a
class of nonlinear covariant gauges},  J.\ Phys.\ A  {\bf 40} (2007)
4025.

\bibitem{Dudal:2005zr}
D.~Dudal, J.~A.~Gracey, V.~E.~R.~Lemes, R.~F.~Sobreiro,
S.~P.~Sorella, R.~Thibes and H.~Verschelde, \emph{Remarks on a class
of renormalizable interpolating gauges}, JHEP {\bf 0507} (2005) 059.

\bibitem{Jackiw:1995nf}
R.~Jackiw and S.~Y.~Pi, \emph{Threshhold Singularities and the
Magnetic Mass in Hot {QCD}}, Phys.\ Lett.\ B {\bf 368} (1996) 131.

\bibitem{Becchi:1975nq}
C.~Becchi, A.~Rouet and R.~Stora, \emph{Renormalization Of Gauge
Theories}, Annals Phys.\  {\bf 98} (1976) 287.

\bibitem{Tyutin}
I.~V.~Tyutin, Lebedev Institute preprint (unpublished), 1975.

\bibitem{Kugo:1979gm}
T.~Kugo and I.~Ojima, \emph{Local Covariant Operator Formalism Of
Nonabelian Gauge Theories And Quark Confinement Problem}, Prog.\
Theor.\ Phys.\ Suppl.\  {\bf 66} (1979) 1.

\bibitem{Slavnov:1989jh}
A.~A.~Slavnov, \emph{Physical Unitarity In The Brst Approach},
Phys.\ Lett.\ B {\bf 217} (1989) 91.

\bibitem{Frolov:1989az}
S.~A.~Frolov and A.~A.~Slavnov, \emph{Construction Of The Effective
Action For General Gauge Theories Via Unitarity}, Nucl.\ Phys.\ B
{\bf 347} (1990) 333.

\bibitem{Peskin:1995ev}
 M.~E.~Peskin and D.~V.~Schroeder, \emph{An Introduction To Quantum Field
Theory}, Addison-Wesley Publishing Company (1995).

\bibitem{Lehmann:1954rq}
 H.~Lehmann, K.~Symanzik and W.~Zimmermann, \emph{On the formulation of quantized field theories}, Nuovo Cim.\  {\bf 1} (1955) 205.

\bibitem{Henneaux:1992ig}
 M.~Henneaux and C.~Teitelboim, \emph{Quantization of gauge systems}, Princeton University Press (1992).

\bibitem{Piguet:1995er}
O.~Piguet and S.~P.~Sorella, \emph{Algebraic renormalization:
Perturbative renormalization, symmetries and anomalies},  Lect.\
Notes Phys.\  {\bf M28} (1995) 1.

\bibitem{Ferrara:1977mv}
 S.~Ferrara and B.~Zumino,
\emph{Structure of linearized supergravity and conformal
supergravity }, Nucl.\ Phys.\ B {\bf 134} (1978) 301.

\bibitem{Ruegg:2003ps}  H.~Ruegg and M.~Ruiz-Altaba,
\emph{The Stueckelberg field}, Int.\ J.\ Mod.\ Phys.\ A \textbf{19}
 (2004) 3265.




\end{thebibliography}
\end{document}